\documentstyle[]{mn}

\newif\ifAMStwofonts

\title[Short timescale variability in the Faint Sky Variability
 Survey]{Short timescale variability in the Faint Sky Variability
 Survey} \author[L. Morales-Rueda et al.]{L. Morales-Rueda$^{1}$,
 P. J. Groot$^{1}$, T. Augusteijn$^{2}$, G. Nelemans$^{1}$,
 P. M. Vreeswijk$^{3,4}$ \newauthor E. J. M. van den Besselaar$^{1}$\\
 $^{1}$IMAPP, Department of Astrophysics, Radboud University Nijmegen,
 P.O. Box 9010, 6500 GL Nijmegen, The Netherlands.\\ E-mail:
 lmr@astro.ru.nl, pgroot@astro.ru.nl, nelemans@astro.ru.nl,
 besselaar@astro.ru.nl\\ $^{2}$Nordic Optical Telescope, Apartado 474,
 E-38700 Santa Cruz de La Palma, Spain. E-mail: tau@not.iac.es\\
 $^{3}$European Southern Observatory, Alonso de C\'{o}rdova 3107,
 Vitacura, Casilla 19001, Santiago 19, Chile. E-mail:
 pvreeswi@eso.org\\ $^{4}$Departamento de Astronom\'{\i}a, Universidad
 de Chile, Casilla 36-D, Santiago, Chile.}
\begin{document}

\date{Accepted ... Received ...; in original form ...}


\maketitle

\label{firstpage}

\begin{abstract}
We present the V band variability analysis of the point sources in the
Faint Sky Variability Survey on time scales from 24 minutes to tens of
days. We find that about one percent of the point sources down to V =
24 are variables. We discuss the variability detection probabilities
for each field depending on field sampling, amplitude and timescale of
the variability. The combination of colour and variability information
allows us to explore the fraction of variable sources for different
spectral types. We find that about 50 percent of the variables show
variability timescales shorter than 6 hours. The total number of
variables is dominated by main sequence sources. The distribution of
variables with spectral type is fairly constant along the main
sequence, with 1 per cent of the sources being variable, except at the
blue end of the main sequence, between spectral types F0--F5, where
the fraction of variable sources increases to about 2 percent. For
bluer sources, above the main sequence, this percentage increases to
about 3.5. We find that the combination of the sampling and the number
of observations allows us to determine the variability timescales and
amplitudes for a maximum of 40 percent of the variables found.  About
a third of the total number of short timescale variables found in the
survey were not detected in either B or/and I. These show a similar
variability timescale distribution to that found for the variables
detected in all three bands.
\end{abstract}

\begin{keywords}
surveys -- methods: data analysis -- stars: general -- stars:
statistics -- stars: variables: general
\end{keywords}

\section{Introduction}

There is a wide range of photometrically variable systems in the
universe. The range of timescales on which these systems vary is as
wide as the physical processes that produce their variability. For
example we have intrinsically variable stars, where the variability is
caused by changes in their internal structure or atmosphere that vary
with timescales of minutes to years \cite{bg94}. Other stars show
variability because they rotate and their surface is inhomogeneous,
e.g. because of star spots, \cite{b05}, or because they form part of a
binary or multiple system and their revolution around the centre of
mass of the system results in changes on the detected flux due to the
changing aspect of a non-isotropically emitting surface or
eclipses. This is also the case for planets orbiting stars. The
timescale of the variability in this case is dictated by the orbital
parameters of the system and can range from seconds to years. Near
Earth Objects (NEOs), such as asteroids, also show variability as they
rotate and are non-spherical. We find photometric variability in
extragalactic objects as well, such as quasars, where the variability
is probably the result of material being accreted by the central
engine, or ``one of'' systems such as gamma ray bursts (GRB) or
supernovae (SNe) where the variability is produced by intrinsic
changes in the structure of an astronomical object that take place
only once.

The study of variability provides important information about the
physical nature of the variable objects, leads to the discovery of new
classes of objects, helps to study the physical structure of stars,
e.g. pulsating stars, allows us to obtain information on galactic
structure through the use of variables such as RR~Lyrae as standard
candles, and is the key to determining extra-galactic distances
through the use of standard candles such as Cepheids and supernovae
Type Ia.

Most of our knowledge of variability is based on the study of
apparently bright sources, which naturally selects members of {\it
intrinsically} bright populations. At present little is known about
variability of intrinsically fainter populations because in bright
samples they are lacking altogether or are only represented by a few
members.  The Faint Sky Variability Survey (FSVS; Groot et
al. \shortcite{groot03}) was designed to account for this deficit by
studying two unexplored regions of the variability space: the short
timescale variability region (down to tens of minutes) and the
intrinsically faint variable sources (down to V = 24 mag) at mid and
high Galactic latitudes.  The FSVS also contains colour information
for all targets, giving us the option of positioning objects in the
colour-colour diagram as well as finding the variability timescales
and amplitudes that characterise them.  The main aims of the FSVS are
thus to obtain a map of a region of the Galaxy ($\sim$21\,deg$^2$) in
variability and colour space, to determine the population density of
the different variable objects that reside in the Galaxy and to find
the photometric signature of up-to-now unknown intrinsically faint
variable populations. In this paper we explore these three goals.

There are other surveys that study the variable optical sky, each
emphasising one aspect or one particular region of this parameter
space. The timescales sampled, depth and sky coverage of different
variability surveys varies depending on the astronomical objects they
are designed to study. For example, with a brightness limit similar to
the FSVS, Street et al. \shortcite{s05} study the variability around
an open cluster with timescales longer than a few hours, and Ramsay
\&\ Hakala \shortcite{rh05} study the rapid variability (down to 2
minutes) of objects as faint as V$\sim$22.5. Of great interest is the
Deep Lens Survey (DLS) that, in a similar way to the FSVS, combines
colour and variability information and explores similar variability
timescales \cite{dls}. Becker et al. \shortcite{dls} also provide a
comprehensive review of past and on-going variability surveys.

The future of optical variability surveys looks quite promising with
the advent of large aperture telescopes such as the Large-aperture
Synoptic Survey 8.4\,m Telescope \cite{tyson02}, the 4\,m telescope
VISTA and the 2.5\,m VLT Survey Telescope.

\section{Observations}

The full Faint Sky Variability Survey (FSVS) data set consists of 78
Wide Field Camera (WFC) fields taken with the Isaac Newton telescope
(INT) at La Palma. The FSVS covers an area in the sky of
$\sim$21\,deg$^2$ located at mid and high Galactic latitudes
($-$40$<b<$$-$21, 15$<b<$50, 89$<b<$90). The WFC is a moisac of four
2kx4k CCDs. For each field, we took one set of B, I and V band
observations on a given photometric night. Photometric variability
observations were taken with the V filter on several consecutive
nights. On average, fields were observed 10 -- 20 times within one
week in the V band. Exposure times were 10 min with a dead time
between observations of 2 min.  This observing pattern allows us to
sample periodicity timescales from 2$\times$(observing time + dead
time) (i.e. 24 min) to twice the maximum time separation of
observations (which ranges from 3 days to 13 days). See for more
details Groot et al. \shortcite{groot03}.

All fields were re-observed years later to determine proper
motions. In this paper we concentrate on the shorter timescale
variability (from 0.4 hours to a few days) of the targets and we do
not include those observations.

\begin{table}
\centering
\begin{minipage}{84mm}
\caption{Journal of observations. The number of V band observations
  taken for each field, not counting those taken about a year later
  for proper motion studies, as well as the maximum time interval
  covered in days, $\Delta t$, are given. For field 48, the two V band
  measurements were taken more than a year apart. For field 51, two
  measurements were lost in one of the CCDs. The fraction of variable
  sources (FV) x100 per field found using the $\chi^2$ test is also
  given. This will be discussed in detail in Sections~\ref{an:chisq}
  and \ref{res:varnovar}.  Notice that the fraction of variables found
  in fields 36, and 48, is very high compared to the rest of the
  fields. These fields were only observed in 2 occasions so their
  variability $\chi^2$ was calculated based only on 2 points. This is
  also the case for fields 35, and 41 to 45 but they do not show a
  fraction of variables as large. These 6 fields contain about twice
  the number of stars compared to fields 36 and 48.}
\label{obs:jo}
\begin{center}
\begin{tabular}{rccl|rccl}
\hline
Field & V obs & $\Delta t$& FV & Field & V obs & $\Delta t$ & FV\\
\hline
01 & 10 & 5.07 & 0.59 & 40 & 9 & 3.07 & 3.44\\
02 & 11 & 5.07 & 1.04 & 41 & 2 & 0.06 & 3.22\\
03 & 13 & 5.11 & 0.21 & 42 & 2 & 0.91 & 3.91\\
04 & 12 & 3.04 & 1.19 & 43 & 2 & 0.03 & 1.62\\
05 & 12 & 3.09 & 0.55 & 44 & 2 & 0.03 & 2.05\\
06 & 10 & 3.11 & 0.74 & 45 & 2 & 0.03 & 1.90\\
07 & 10 & 5.99 & 0.48 & 46 & 19 & 4.18 & 1.06\\
08 & 11 & 5.07 & 0.09 & 47 & 5 & 6.97 & 4.20\\
09 & 10 & 5.02 & 0.79 & 48 & 2 & - & 8.60\\
10 & 10 & 3.02 & 0.09 & 49 & 3 & 5.94 & 5.26\\
11 & 7 & 3.04 & 0.23 & 50 & 3 & 6.07 & 3.75\\
12 & 8 & 3.07 & 0.31 & 51 & 5,3 & 5.92 & 2.09\\
13 & 9 & 5.06 & 0.77 & 52 & 26 & 5.17 & 0.80\\
14 & 10 & 5.03 & 0.54 & 53 & 20 & 6.10 & 0.24\\
15 & 8 & 3.03 & 0.55 & 54 & 20 & 6.01 & 0.15\\
16 & 12 & 3.06 & 0.51 & 55 & 20 & 5.98 & 1.11\\
17 & 8 & 3.09 & 0.64 & 56 & 19 & 4.18 & 0.37\\
18 & 9 & 5.06 & 0.67 & 57 & 18 & 5.24 & 1.30\\
19 & 11 & 4.02 & 0.35 & 58 & 15 & 4.18 & 0.70\\
20 & 11 & 4.04 & 0.11 & 59 & 20 & 5.12 & 0.60\\
21 & 11 & 4.07 & 0.74 & 60 & 22 & 4.16 & 1.93\\
22 & 11 & 4.05 & 2.73 & 61 & 20 & 4.09 & 0.47\\
23 & 11 & 4.08 & 3.06 & 62 & 19 & 4.06 & 1.95\\
24 & 10 & 4.11 & 1.62 & 63 & 22 & 6.09 & 0.52\\
25 & 30 & 6.12 & 0.06 & 64 & 22 & 6.06 & 0.43\\
26 & 29 & 5.00 & 0.58 & 65 & 20 & 6.03 & 0.40\\
27 & 14 & 6.01 & 0.25 & 66 & 21 & 6.00 & 0.41\\
28 & 14 & 5.99 & 0.66 & 68 & 28 & 12.07 & 1.11\\
29 & 14 & 6.02 & 0.45 & 69 & 29 & 13.00 & 1.50\\
30 & 16 & 4.99 & 1.60 & 70 & 24 & 6.03 & 0.88\\
31 & 16 & 4.99 & 0.87 & 71 & 27 & 6.00 & 1.04\\
32 & 16 & 4.99 & 0.36 & 72 & 26 & 6.11 & 0.50\\
33 & 18 & 6.10 & 0.18 & 73 & 27 & 6.09 & 0.66\\
34 & 17 & 6.07 & 0.42 & 74 & 26 & 6.05 & 0.23\\
35 & 2 & 0.03 & 1.83 & 75 & 25 & 6.00 & 0.21\\
36 & 2 & 0.97 & 13.68 & 76 & 33 & 7.03 & 0.37\\
37 & 10 & 4.08 & 0.35 & 77 & 33 & 7.06 & 0.48\\
38 & 9 & 4.05 & 0.73 & 78 & 33 & 7.15 & 0.48\\
39 & 9 & 3.09 & 0.77 & 79 & 31 & 7.15 & 0.28\\
\hline
\end{tabular}
\end{center}
\end{minipage}
\end{table}

Of the 78 fields, 10 were only observed in the V band on two or three
occasions due to bad weather, making it impossible to use these for
precise variability studies. We also encountered data
acquisition/reduction problems in several occasions which resulted in
parts of fields being lost.  Initially 79 fields were defined but one,
67, was never observed. For data handling consistency we have kept the
original numbering.

A complimentary analysis on the variability at short and long
timescales (including year-long timescales) of the FSVS has been
carried out by Huber et al. \shortcite{h06}. They make use of a
variability test similar to that described in Section~\ref{an:chisq}
to find signatures of the presence of variability with some indication
of its timescale and amplitude. By using the yearly re-observations
they find a variability fraction of 5 to 8 percent in two survey
regions. This number is larger than that found in
Section~\ref{res:varnovar} due to several factors: in our analysis we
are only considering short period variables and thus do not make use
of the yearly re-observations, we determine the variability fraction
using the entire area of the survey instead of two separate regions,
and we use a reduced weighted $\chi^2$ to establish variability
instead of the reduced un-weighted $\chi^2$ found in the public
release FSVS data products. Most of the variable sources found by
Huber et al. \shortcite{h06} are long period variables classified as
such thanks to the V brightness of the re-observation a year
later. The number of possible periodicities in the data when the
sampling is sparse and the time span long is very large and their
analysis is devised to find possible variable systems more than to
find their actual variability timescale and amplitude, which is one of
the main goals of the work presented here.

Table~\ref{obs:jo} gives a list of the number of times each field was
observed in V and the maximum time span of the data in each case.  In
summary, we have photometric data that can be used for variability
analysis for $\sim$17.5\,deg$^2$ out of the 21\,deg$^2$ that
constitute the FSVS. For $\sim$9.2\,deg$^2$ the number of measurements
is equal or more than 15 whereas for the other $\sim$8.3\,deg$^2$ the
number is between 5 and 15. This makes a difference in the accuracy
with which we can measure the variability timescale of each object.

The methods used to study the variability in the data are presented in
Section~3. The results from the variability analysis are discussed in
the different subsections of Section~4. In this paper we carry out a
full variability study for point sources that have not only more than
4 V band measurements taken over a two week baseline, but also
positive detections in the B and I bands. Possible extreme colour
systems are discussed in detail in Section~\ref{res:extremecolour}.

\subsection{Data quality checks}
\label{obs:quality}

We carried out several tests to check the quality of the
photometry. These included plotting several quantities obtained from
the data to check for anomalies. We explored how the number of
detected point sources changed with epoch for each field, the average
point source V magnitude per epoch per field and the ratio between the
point source mean magnitude and the limiting magnitude for each
measurement.

\begin{figure*}
\begin{picture}(100,0)(-270,250)
\put(-300,230){\includegraphics{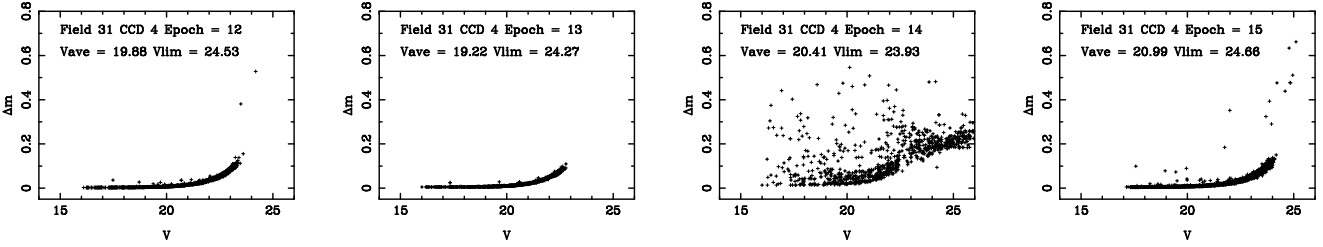}}
\noindent
\end{picture}
\vspace{35mm}
\caption{V magnitude versus V error for each point source detected in
  Field 31 CCD 4 in 4 different epochs. These plots were generated for
  all fields to check the quality of the data. See text for details.}
\label{obs:quality:fig1}
\end{figure*}

We identified several fields that showed anomalies, such as field 31
CCD 4, in which one of the observations (number 14) resulted in V band
magnitudes that were lower than the rest by 2 magnitudes. We found
that the best way to identify these anomalies consisted of plotting
the V band magnitudes for each point source detected in each field
versus its error, for each observation. An example of this test is
shown in Fig.~\ref{obs:quality:fig1} where the point sources found in
Field 31 CCD 4 on four different epochs are plotted. Each panel
presents the test for one observation. The average V band magnitude
(Vave) for all point sources and the limiting magnitude (Vlim) for
each image is also given. When the field was observed through thin
clouds (e.g. epoch 13) the values of Vlim and Vave decrease but the
shape of the curve does not change. On the other hand if something
went wrong with the image or the data reduction we expect the shape of
the curve to change (e.g. epoch 14) giving us an indication that we
should be wary of this V point when doing our variability searches. We
could not trace the reason for the anomaly found in epoch 14 of field
31 and just discarded this data point.

We also carried out visual inspection of the raw unfolded lightcurves
for all the variable point sources to identify possible problematic
photometry points and when confirmed these points were thrown out.

\section{Variability analysis methods}
\subsection{The variability $\chi^2$ test}
\label{an:chisq}

Groot et al. \shortcite{groot03} determine the variability of a given
point source in the FSVS by calculating the reduced $\chi^2$ value of
the object's individual brightness measurements with respect to its
weighted mean brightness value. An object is tagged as variable if its
reduced $\chi^2$ is above the 5-$\sigma$ level. This definition will
be used in Sect.~\ref{res:varnovar} to determine the fraction of
variable objects in the survey. Because we have colour information for
the majority of the objects, we can determine this ratio for different
types of systems.

\subsection{The floating mean periodogram}
\label{an:fmp}

If we not only want to know whether an object is variable or not, but
also what the timescale and amplitude of its variability are, and thus
what type of object it might be, we need to use more refined methods
to determine its variability. Because of the relatively small number
of V band observations (between 2 and 33 depending on the field) we
use the ``floating mean'' periodogram technique to estimate the
characteristic variability timescale in each case. This method works
better than the traditional Lomb-Scargle algorithm
\cite{lomb76,scargle82} for small number of points and has been
successfully applied to planet searches \cite{cumming99} and to
determine the orbits of subdwarf B binaries \cite{moralesrueda03}. A
minimum number of 5 V measurements is required to calculate the
floating mean periodogram.

The floating mean periodogram consists of fitting the data with a
model composed of a sinusoid plus a constant of the form:
\begin{displaymath}
A(t) = \gamma + K \sin(2 \pi(t - t_0)/P),
\end{displaymath}

where $\gamma$ is the average V magnitude, $K$ is the amplitude of the
V variability, $P$ is the period and $t$ is the time of
observation. For each given period we perform singular value
decomposition least square fitting of the data solving for $\gamma$
and $K$ \cite{press92}. We obtain the $\chi^2$ of the fits as a
function of frequency $f$ = $1/P$ and select the minima of this
$\chi^2$ function.

\begin{figure*}
\begin{picture}(100,0)(-270,250)
\put(-300,250){\includegraphics{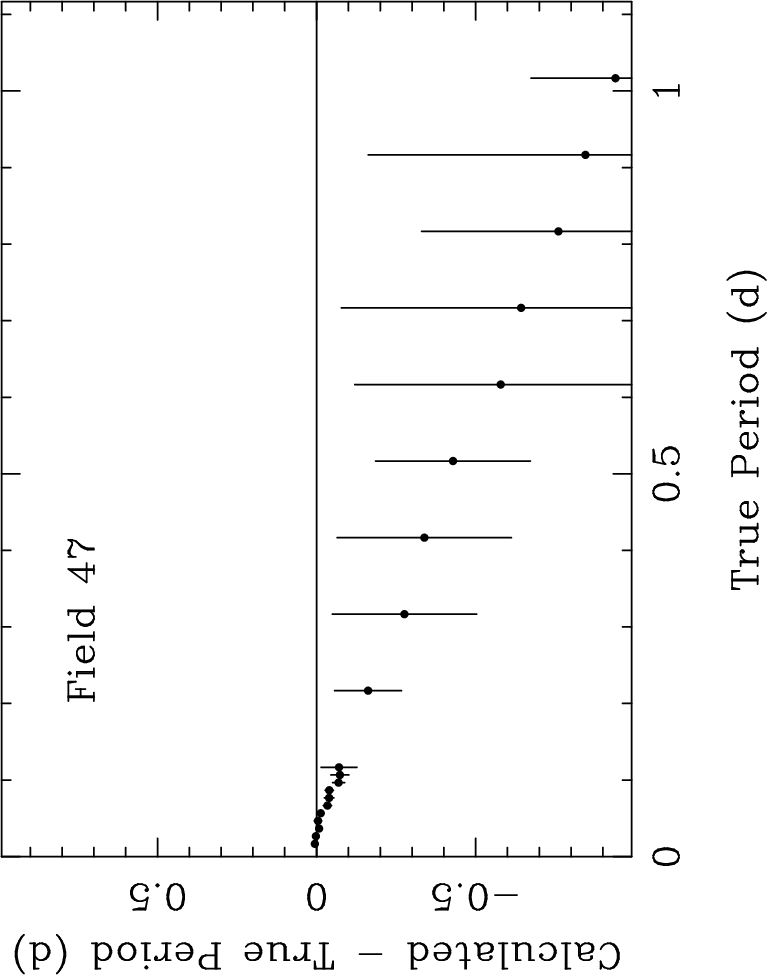}}
\put(-300,250){\includegraphics{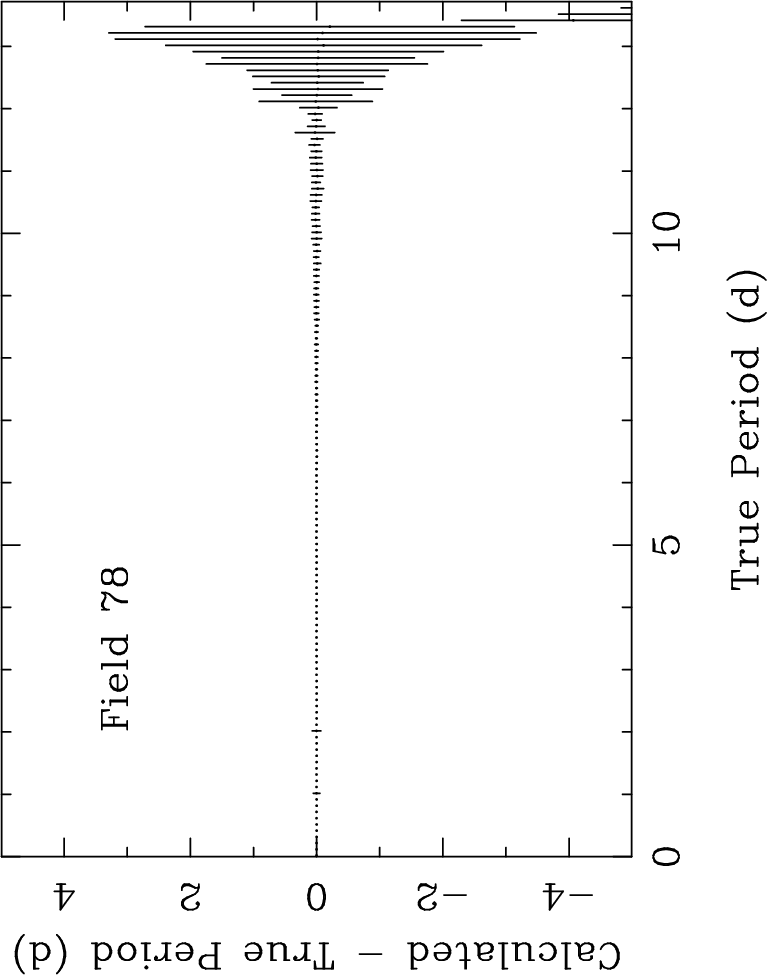}}
\noindent
\end{picture}
\vspace{65mm}
\caption{True period versus calculated-true period using the floating
  mean periodogram for two example fields, 47 with 5 V band
  measurements and 78 with 33 measurements. The period search was
  stopped when either the difference between the calculated and the
  true period was larger than 3$\times$ the error in the period, or
  the maximum searchable period was reached, i.e. 2$\times$ time
  baseline (although this last condition was never reached). The ratio
  of the amplitude of the variability to error assumed in both
  examples is 50.}
\label{an:fmp:fig1}
\end{figure*}

To test whether the periodogram was able to recover the correct
periods we generated, for at least 9 different amplitudes per field,
synthetic lightcurves that vary sinusoidally with periods between 24
min and several days (we chose the upper limit to be twice the time
span of the observations) in period steps of 0.1 days, using the time
sampling of each field. For each amplitude and period, we generated
lightcurves with 20 different phasings, i.e. 20 different t$_0$. For
each of these 20 phasings, we generated 10 lightcurves where the
magnitude for each point was calculated given the period, the phasing,
the amplitude and a fixed average magnitude. An error was added to the
resulting magnitude for each point. This error was calculated by
drawing a random number from a normal distribution centred on zero
with standard deviation of 0.03, which is the average V band error
found in the FSVS data.

We then used the floating mean periodogram to calculate the most
probable period for each lightcurve, averaging the periods obtained
for the 200 different lightcurves generated for each input timescale
and amplitude. The average obtained was a weighted average where the
weights used were the errors of the periods determined. We stopped the
period search when the difference between the true and the calculated
period was larger than 3$\sigma$. This condition was always reached
before we got to the maximum period allowed in the search. We noticed
large deviations (although still within the 3$\sigma$ difference)
between the calculated and true values at certain isolated periods.
In these cases the true value for the following periods was again
successfully recovered. These deviations are caused by the sampling
windows of each given field. The presence of these isolated deviations
prompted us to select the criteria described above to stop the
simulated lightcurve fitting even when, in occasions, the errors on
the calculated period were so large as to render the period
determination highly inaccurate. To account for this inaccuracy, we
will apply further filtering criteria to the errors of the periods
determined from the data in Section~\ref{res:timeamp}.

\begin{figure*}
\begin{picture}(100,0)(-270,250)
\put(-300,250){\includegraphics{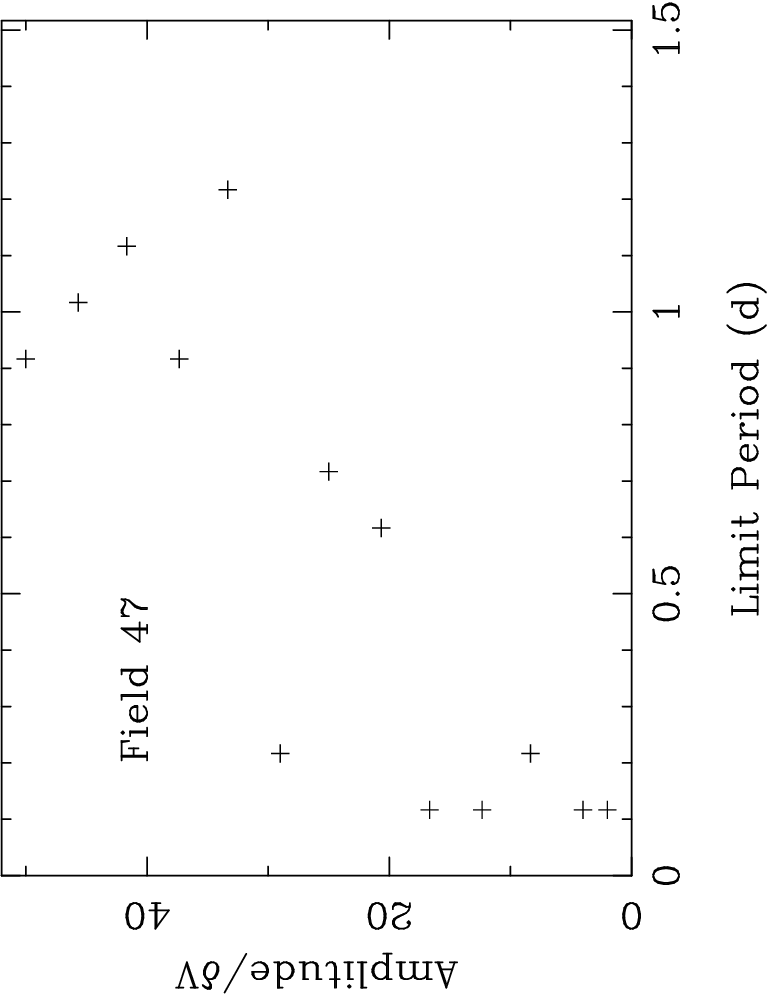}}
\put(-300,250){\includegraphics{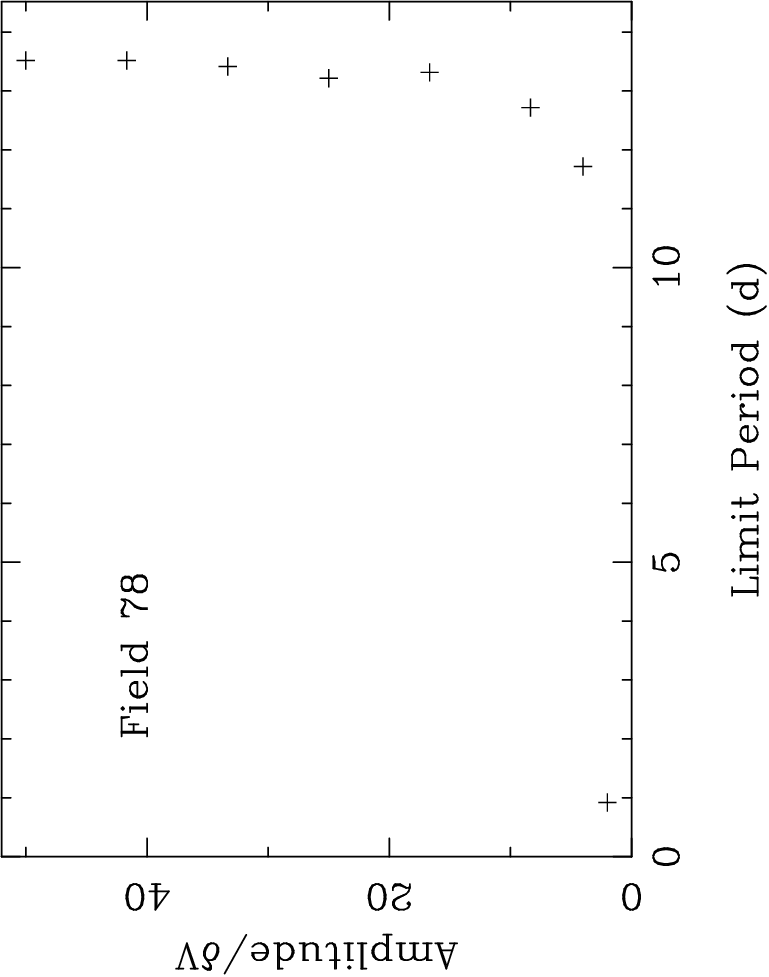}}
\noindent
\end{picture}
\vspace{65mm}
\caption{Maximum period that can be reconstructed from the simulated
  data depending on the ratio of the amplitude of the variability and
  the error in V ($\delta$V). The period search was stopped when the
  same conditions described in Fig.\ref{an:fmp:fig1} were reached. The
  period search was carried out fixing the error in V to 0.03 mags and
  changing the amplitude of the variability. This is the average error
  in brightness we find in the FSVS.}
\label{an:fmp:fig2}
\end{figure*}

After carrying out the simulations, the result for each field, or at
least for the ones with more than 10 observations, is that true
periods are recovered successfully with an error that increases with
period. A few fields also show small deviations (but still within
3$\sigma$) from the true values at the shortest period (24 min). A
common trend we observe is that, as we are reaching the limit period,
the calculated values tend to underestimate the true periods at the
same time that the error in the calculated period increases. This is
clear in both panels of Fig.~\ref{an:fmp:fig1} where we have plotted
the calculated period versus the true period. This seems to indicate
that, when pushed to the limit, the periods calculated by the floating
mean periodogram will be shorter than the true
ones. Fig.~\ref{an:fmp:fig1} shows, as an illustration, two example
graphs for two fields that have a very different number of
observations, field 47 with only 5 V band observations and field 78
with 33 observations. Notice that not only the number of observations
is of importance, but that using a different time sampling also
generates slightly different graphs even with the same number of
observations. The amplitude of the variability assumed also influences
indirectly the maximum period we can determine in each
case. Specifically, it is the ratio between the amplitude of the
variability and the errors in the photometry which influence the
success in recovering a period. By fixing the photometric errors in
the simulations to $\sigma_{\rm average}$ = 0.03 mag, we explore the
influence of this ratio by varying only the amplitude of the
variability. The graphs in Fig.~\ref{an:fmp:fig1} have been obtained
assuming a variability amplitude of 1.5 mags (the maximum value used
in the simulations) equivalent to an amplitude - V band error ratio of
50.

We obtain similar curves for all the fields to determine how effective
the algorithm is at finding the true periods depending not only on the
sampling and the number of observations, but also on the amplitude of
the variability and the brightness of the objects (which define the
errors in the V band magnitudes). For each field we generate a curve
like those shown in Fig.~\ref{an:fmp:fig2}, which present the maximum
period that we can reconstruct from the simulated data for a range of
variability amplitudes. As expected, the larger the amplitude of the
variability (equivalent to a larger amplitude-error ratio), the longer
the timescale of the variability that we can detect. The sampling and
the time span of the observations have a direct influence on the
maximum period we can detect. The graphs are presented only for the
two example fields, 47 and 78, where the number of V band observations
is very different over the same time span, $\sim$7\,d. In the case of
field 47, with only 5 measurements, the maximum period we can
reconstruct before the difference between true and calculated period
is larger than 3$\sigma$ is about 0.4\,d (see left panel of
Fig.~\ref{an:fmp:fig1}). Its limit period graph (left panel of
Fig.~\ref{an:fmp:fig2}) shows some departures from the expected
behaviour, i.e. higher limit period as the amplitude/error ratio
increases. In contrast, for field 78, with 33 V measurements we can
detect periods of up to 13.5\,d for the same variability amplitude and
the limit period graph shows the expected behaviour. A similar number
of measurements over a shorter time span will only allow us to measure
shorter variability timescales. E.g. for field 26, with 29
measurements over 5\,d, the maximum period we can reconstruct (for
variability amplitude of 1.5 mags) is 9.3\,d.

We want to remind the reader at this point, that by using the floating
mean periodogram we are fitting the data with sinusoidal curves. Any
non-sinusoidal, or indeed any non-periodic variability present in the
data will be poorly fitted.

\begin{figure*}
\begin{picture}(100,0)(-270,250)
\put(-300,230){\includegraphics{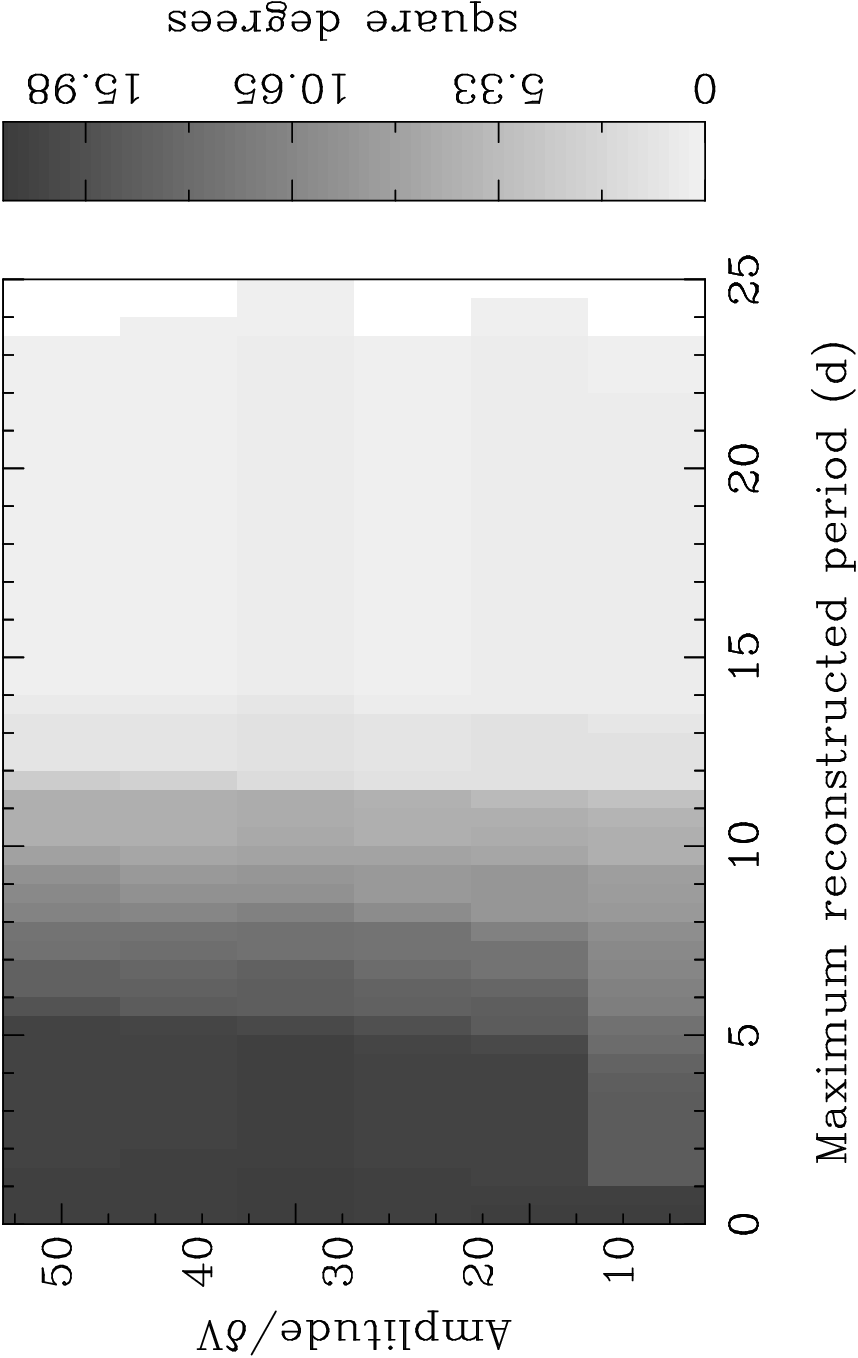}}
\put(-300,230){\includegraphics{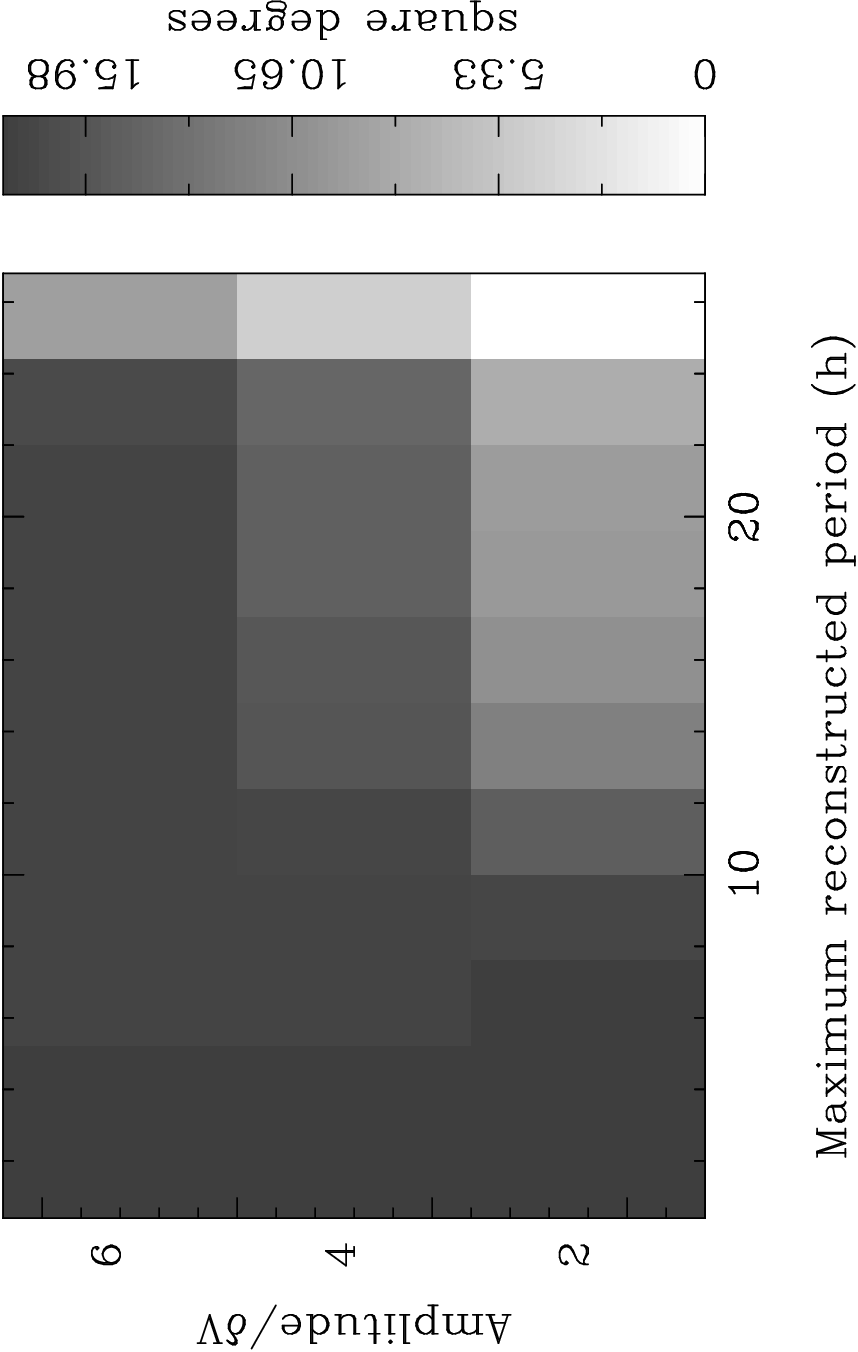}}
\noindent
\end{picture}
\vspace{60mm}
\caption{Left panel: gray scale map presenting the number of $deg^2$
  in the sky for which we can reconstruct a given variability
  timescale depending on the ratio of the variability amplitude and
  the magnitude error. Note that if a variability scale of 5 days can
  be reconstructed for a given field also variabilities smaller than 5
  days can be reconstructed. The minimum limit being the minimum
  possible searchable period of 24\,min.  Right panel: same
  information but zooming into the short period, small amplitude/error
  region. We reconstruct the minimum searchable period for all
  fields. Note that the structure seen on this panel cannot be seen on
  the left hand side panel due to the coarser binning in
  Amplitude/$\delta$V of the left hand side panel.}
\label{an:fmp:fig3}
\end{figure*}

The maximum period that can be reconstructed for all the fields in the
FSVS simultaneously is plotted in Fig.~\ref{an:fmp:fig3}.  For an
amplitude of the variability of $\sim$0.25 mags (equivalent to an
amplitude to magnitude error ratio of $\sim$8.3), we can reconstruct
variability periods of up to 1\,d for $\sim$17.58\,deg$^2$ out of the
18.11\,deg$^2$ available for search (66 fields out of the 68 with more
than 4 V band measurements), we can reconstruct variabilities of up to
5\,d for $\sim$13.31\,deg$^2$ (50 fields). For amplitudes of 1.5 mag
(magnitude error ratio of 50) we can reconstruct periods of up to 11
days for a region of 6.66\,deg$^2$ (25 fields) and so on. We can only
search for periods of the order of 20 days in 2 fields.
 
We are also interested in detecting small amplitude and short period
variability, the limits of the data being a 24\,min sampling and the
photometric accuracy of $\sim$3 millimags for the brightest objects
\cite{groot03}. The right panel of Fig.~\ref{an:fmp:fig3} zooms into
the the short timescale, small amplitude variability region. We
reconstruct successfully (i.e. with less than a 3$\sigma$ difference
between the true and calculated period) the minimum searchable period
in all fields. 

\subsubsection{Our efficiency to detect the variability timescales of some
  interesting astronomical objects}

Some interesting astronomical objects such as Cataclysmic Variables
(CVs) and RR Lyr show characteristic variability timescales. CV
periods range from 80\,min to $\sim$6\,h (although some of them show
longer orbital periods such as GK~Per with a 2\,d orbit).  The
variability of RR~Lyrae ranges from $\sim$6\,h to about 1\,d. Other
interesting systems such as AM~CVn binaries show orbital periods of
the order of tens of minutes, too short to be detected in the FSVS. On
the other hand, orbital periods of 80\,min will be detected in the
full area of the FSVS, as long as the ratio of the amplitude of the
variability and the error in the V magnitude is at least 10, i.e. we
would be able to detect all 80\,min variables down to V = 24 if their
variability amplitude is at least 2 mag, down to V = 23 if the
variability amplitude is at least 0.7 mag, and down to V = 22 if the
variability amplitude is at least 0.36 mag. CVs show characteristic
orbital variability amplitudes of the order of 0.1--0.4 mag thus we
will be able to detect a fraction at least down to V = 22. For certain
fields, when looking at the short period region, the calculated period
underestimates the value of the true period. This will most probably
happen also for the real lightcurves.

Orbital periods of up to 6 hours, and between 6 hours and 1 day (this
last period range is typical of RR~Lyr) will be detected in
17.58\,deg$^2$ (all fields but two) as long as the ratio of the
amplitude of the variability and the V error is at least 20, i.e. we
would be able to detect all variables with periods of up to 1 day in
this area down to V = 24 if the variability amplitude is larger than 4
mag, and down to V = 23 if the variability amplitude is at least 1.4
mag. The variability amplitudes typical of RR~Lyr range between
$\sim$0.5 and $>$1 mag, which indicates that we are sensitive to
RR~Lyr down to V = 23 as long as the variability amplitude is at least
$\sim$1.4 mag.

Other pulsating stars such as $\gamma$\,Doradus stars, $\delta$\,Scuti
stars, slowly pulsating B stars, $\beta$\,Cep stars and short period
Cepheids show pulsation periods and amplitudes in the detectable range
of this survey. Some of them, like $\delta$\,Scuti stars, show very
complicated oscillation patterns that are far from sinusoidal which
means that, although they would be detected as variables with the
$\chi^2$ test, the periods reconstructed with the floating mean
periodogram will most probably be incorrect. Short period pulsators
such as rapidly oscillating Ap stars, PG1159 stars, pulsating subdwarf
B stars and pulsating white dwarfs, and long period pulsators such as
RV Tauri stars and Mira stars will not be detected in the FSVS as
their pulsating periods lie outside the range we are sensitive
to. Solar-like stars show very small amplitude pulsations that cannot
be detected in the FSVS.

Asteroids show rotational periods of the order of a few hours and also
lie in the detectable range of the FSVS. We see a number of asteriod
tracks in the FSVS images but these asteroids do not stay in the same
position from image to image and are discarded during data reduction.

We will use these results again in Section~\ref{res:timeamp} to
estimate how reliable our detections and non-detections for
variability are at given timescales and amplitudes.

\section{Results}

\subsection{Fraction of variable sources}
\label{res:varnovar}

\begin{figure*}
\begin{picture}(100,0)(-270,250)
\put(-310,240){\includegraphics{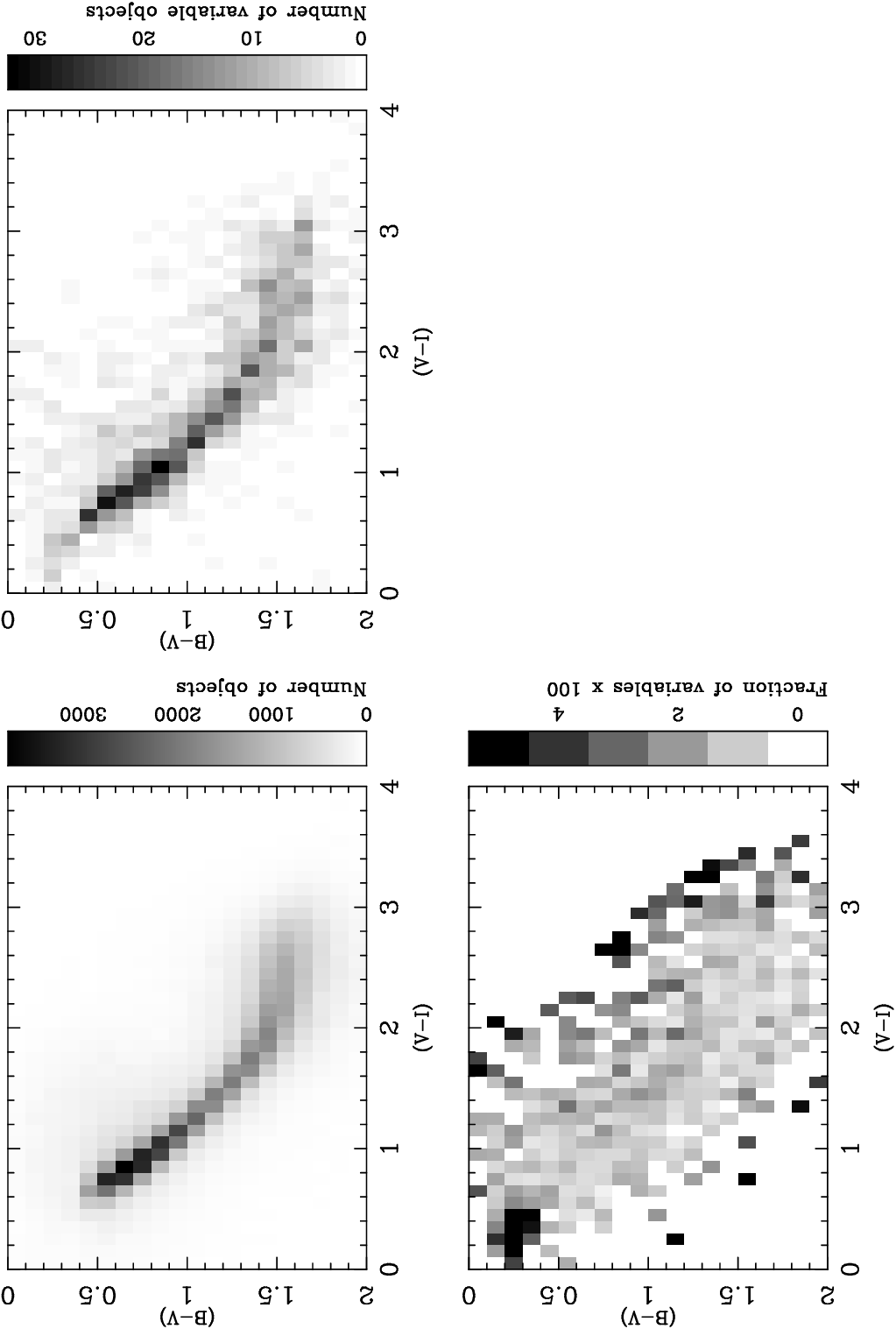}}
\put(-310,240){\includegraphics{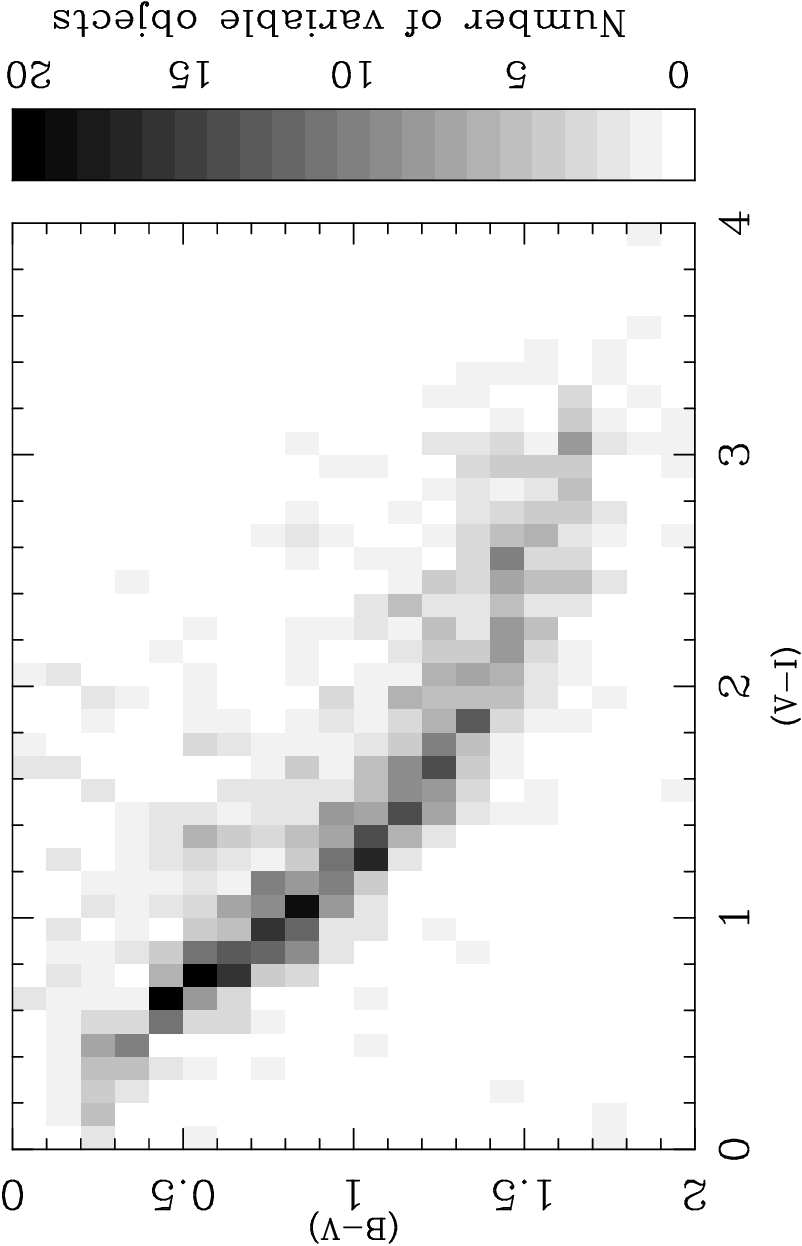}}
\noindent
\end{picture}
\vspace{125mm}
\caption{Top left panel: non variable sources in the FSVS. Top right
  panel: short term variable sources in the FSVS. Bottom left panel:
  fraction of short term variable sources in the FSVS. The fraction is
  presented in percentages. The region in the colour-colour diagram
  where there are more variable sources lies above the blue cut-off,
  i.e.  (B$-$V)$<$0.38, and 0$<$(V$-$I)$<$1. Bottom right panel:
  colour-colour diagram presenting the 976 variable point sources that
  have more than 4 measurements and can be studied using the floating
  mean periodogram method. Note that some of the blue variables found
  in this diagram might be QSOs.}
\label{res:varnovar:fig1}
\end{figure*}

In the entire FSVS, using the $\chi^2$ test, after discarding
problematic points, we find a total of 1\,713 short timescale variable
V-band point sources that have been detected also in the B and I
bands. The number of non variable sources found (after applying the
same criteria as for the variables sources, i.e. account for
problematic points and positive detections in B and I) is 173\,276
($\sim$1 percent of all point sources detected are short term
variables).

In the top left panel of Fig.~\ref{res:varnovar:fig1} we present the
distribution of sources in the FSVS in the form of a grey-scale plot,
which shows that most objects fall along the main sequence with the
largest numbers at its blue end. The top right panel presents the same
plot for the variable sources in the FSVS as determined by the
$\chi^2$ test. Again, most sources are situated along the main
sequence, but there is also a significant fraction of sources above
the main sequence and as a `blue' extension of the main-sequence. To
highlight the difference in distribution between the variable and non
variable sources we present the fraction of variable sources as a
function of colour in the lower left panel (which is basically the
ratio of the upper two panels in Fig.~\ref{res:varnovar:fig1}). This
shows a clear enhancement of sources above the blue tip of the main
sequence (about 4 percent are variable) and a less marked enhancement
to the right and above the main sequence while the main sequence
itself appears as an area with relatively few variables, of the order
of 1 per cent. These values agree with those found by Everett et al.
\shortcite{e02} in a similar, albeit shallower, study.

The fraction of variable point sources found in each field using the
$\chi^2$ test is given in Table~\ref{obs:jo}. The fraction of
variables differs from field to field ranging from 0.06 to 13.68 per
cent. We find that fields with less than 4 measurements tend to have
larger fractions of variables but we are not certain that there is a
direct correlation with the number of observations and that other
effects, such as position in the sky, are not taking place. If we do
not consider the fields with less than 4 observations, we determine a
0.7 per cent fraction of variables in the full FSVS. The distribution
along the main sequence decreases only slightly, by 0.2 per cent, for
all colour-colour bins.

\subsection{Sample of variables studied with the floating mean periodogram}

Ten fields have less than 5 V measurements. The timescale of the
variability for the variables in these 10 fields cannot be determined
because of the low number of measurements, thus only a fraction of the
1\,713 short term variables found in the FSVS can be studied in more
detail using the floating mean periodogram. Once we account for fields
with less than 5 measurements, for problematic epochs, and for objects
that although observed more than 4 times, were only detected 4 times
or less, we end up with 976 point sources that can be studied in
further detail. In most cases the non detection of a source was the
result of faint objects, occasionally falling below the limit of
detection, and in some cases of the objects being blended or
truncated. After accounting for these three factors we find that the
fraction of variable systems found is independent of the number of V
measurements. For the objects with more than 4 detections, information
about the timescale and amplitude of the variability can be
obtained. The distribution of these 976 sources in the colour-colour
diagram is given in the bottom right panel of
Fig.~\ref{res:varnovar:fig1}.

\subsection{Timescale and Amplitude of the variability}
\label{res:timeamp}

When we run the floating mean periodogram on the real data lightcurves
we obtain their most likely variability timescale and the amplitude of
the variability on that timescale. After rejecting those sources for
which the period and amplitude measured lie outside the ranges that
can be reconstructed, according to the simulations carried out in
Section~\ref{an:fmp}, we are left with 744 variable point sources out
of the 976 mentioned above.

To understand this sample of variables, we select different cutoffs
for the error in the variability timescale and amplitude calculated.
Table~\ref{res:timeamp:tab1} gives the fraction of variable point
sources found for different combinations of error cutoffs in timescale
and amplitude. Applying an error cutoff of 30 percent in the periods
and 50 percent in the amplitudes we can already reconstruct the 744
initial variables. Applying an error cutoff of 30 percent in both
period and amplitude we find that 50 percent of the variables show
periods between 24\,min and 6\,hours, 22 percent between 6\,hours and
1\,day, 20 percent between 1 and 4\,days, and 8 percent show periods
above 4\,days. If we apply fairly strict error cutoffs for the period
and amplitudes, i.e. 10 percent, the number of variable sources
decreases to 219, of which the distribution in the same period bins is
51, 20, 19 and 10 percent respectively.

If we assume a normal distribution for the data, a 50 percent error
corresponds to 2$\sigma$, a 30 percent error to 3.3$\sigma$, a 20
percent to 5$\sigma$, and a 10 percent to 10$\sigma$.

\begin{table}
\centering
\begin{minipage}{84mm}
\caption{Fraction of variable objects (out of the 744) found in the
  FSVS for different cutoff limits on the error of the timescale and
  the error of the amplitude.}
\label{res:timeamp:tab1}
\begin{center}
\begin{tabular}{lcccc}
\hline
$\delta$P (\%)    & 10 & 20 & 30 & 40 \\
\hline
$\delta$A (\%) &    &    &    &    \\
10                     & 29.4& 29.7& 29.7& 29.7\\
20                     & 75.7& 76.2& 76.3& 76.3\\
30                     & 91.4& 92.2& 92.6& 92.6\\
40                     & 96.6& 97.4& 97.8& 97.8\\
\hline
\end{tabular}
\end{center}
\end{minipage}
\end{table}

\begin{figure}
\begin{picture}(100,0)(-270,250)
\put(-310,240){\includegraphics{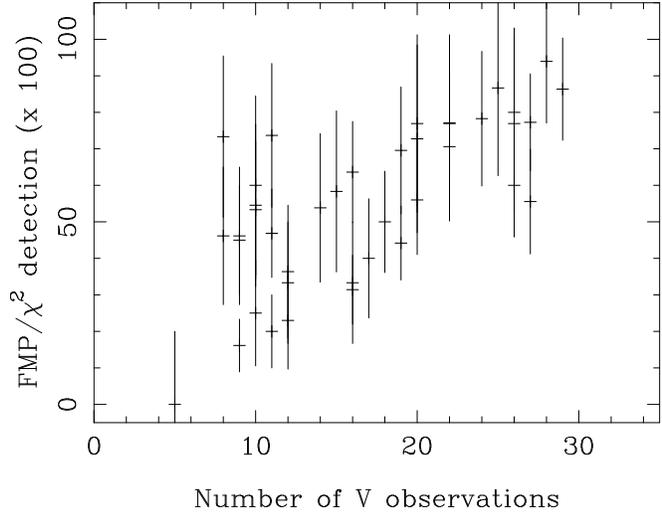}}
\noindent
\end{picture}
\vspace{70mm}
\caption{Ratio of variable sources solved with the floating mean
periodogram with respect to the number of sources flagged as variable
with the $\chi^2$ test, as a function of number of observations. We
have only included the sources determined using the floating mean
periodogram that have errors of less than 20 percent. We only plot
those fields that contain more than 10 variables. FMP stands for
floating mean periodogram.}
\label{res:timeamp:fig1}
\end{figure}

\begin{figure*}
\begin{picture}(100,0)(-270,250)
\put(-310,240){\includegraphics{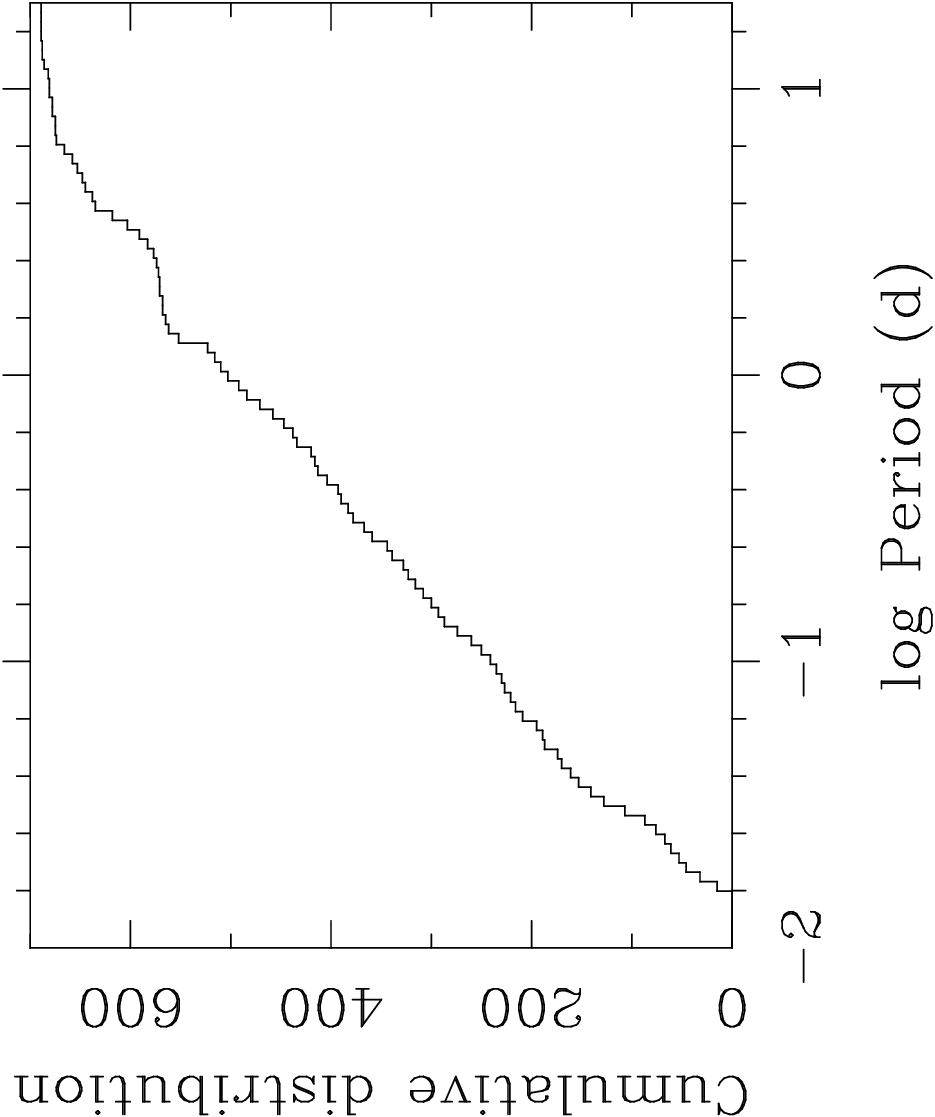}}
\put(-310,240){\includegraphics{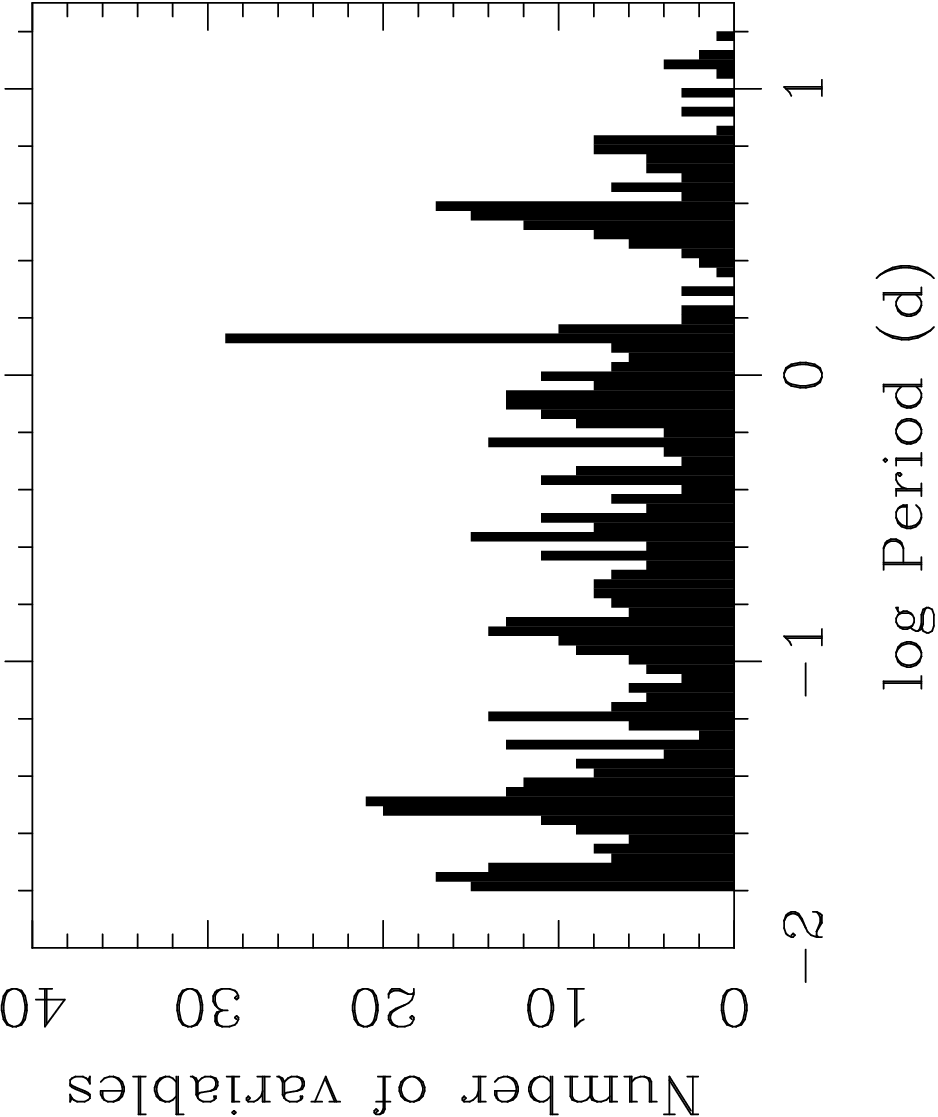}}
\put(-310,240){\includegraphics{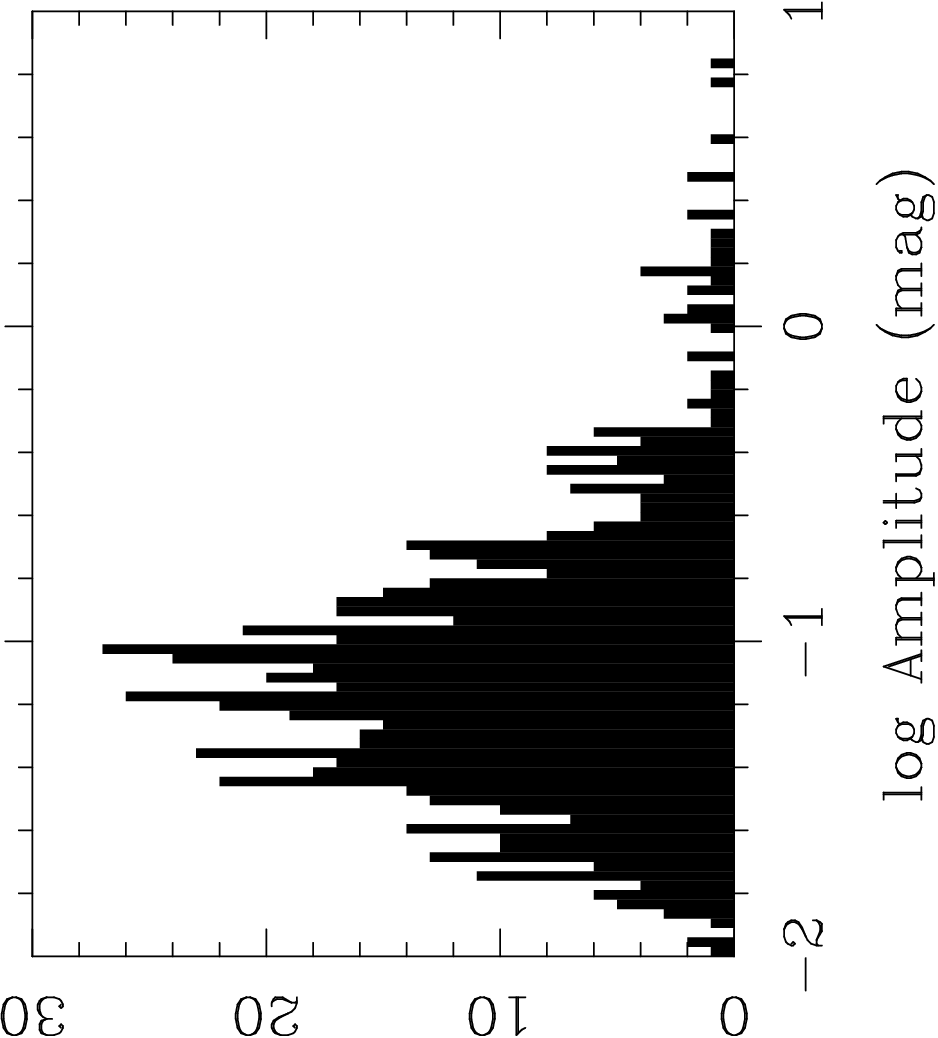}}
\put(-310,240){\includegraphics{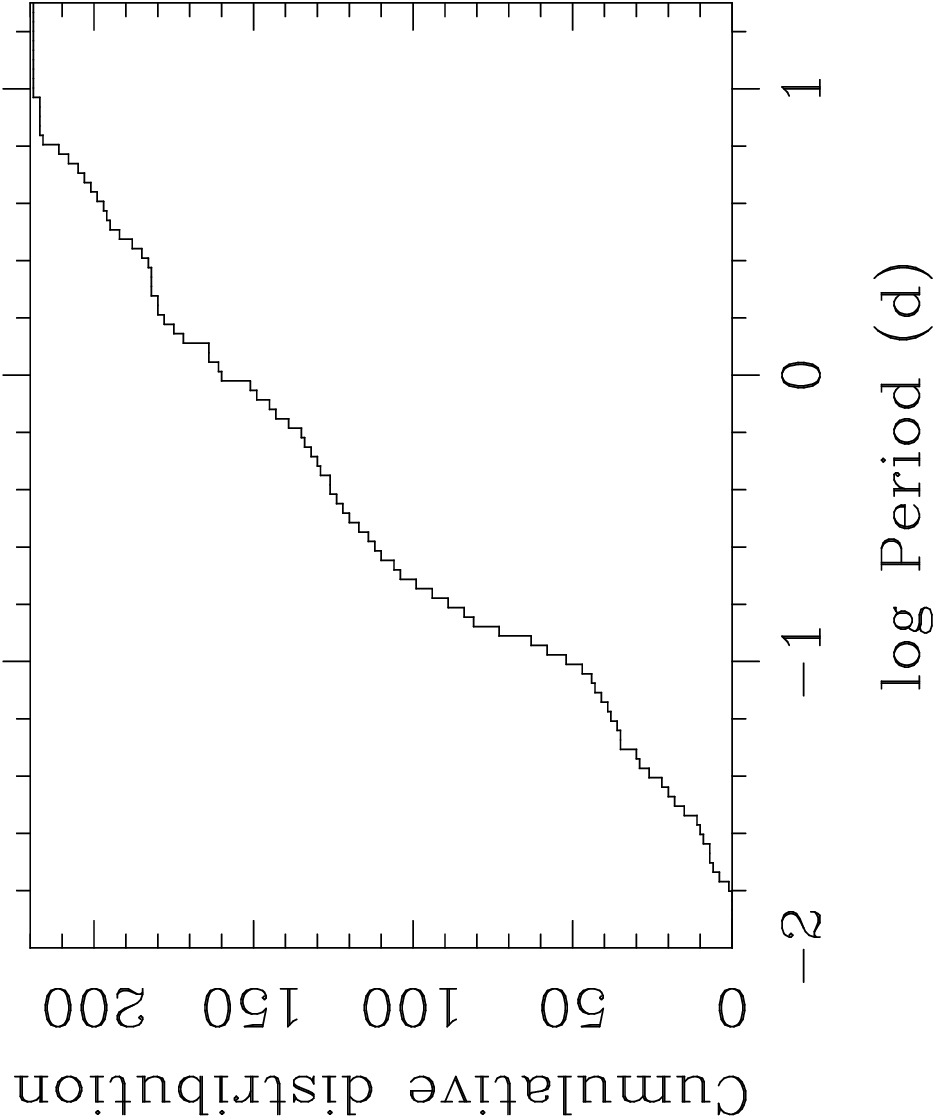}}
\put(-310,240){\includegraphics{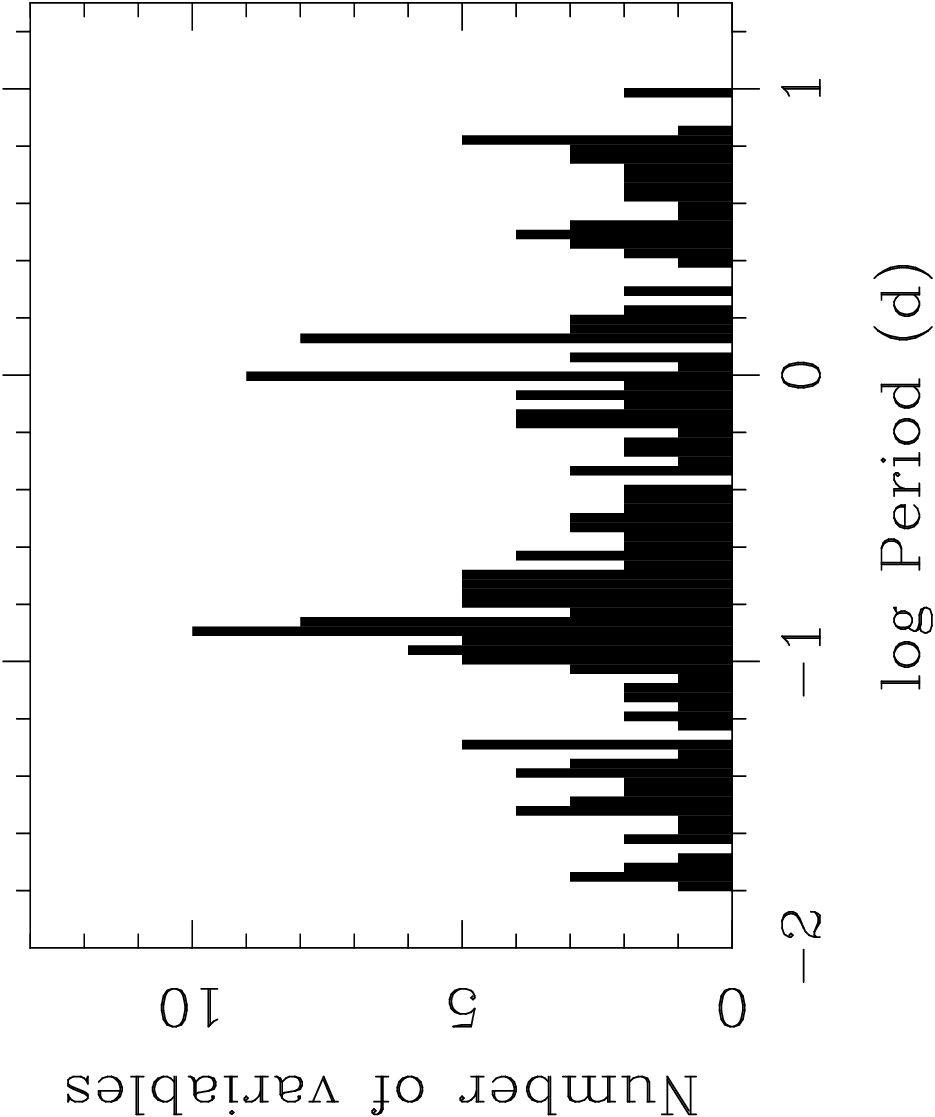}}
\put(-310,240){\includegraphics{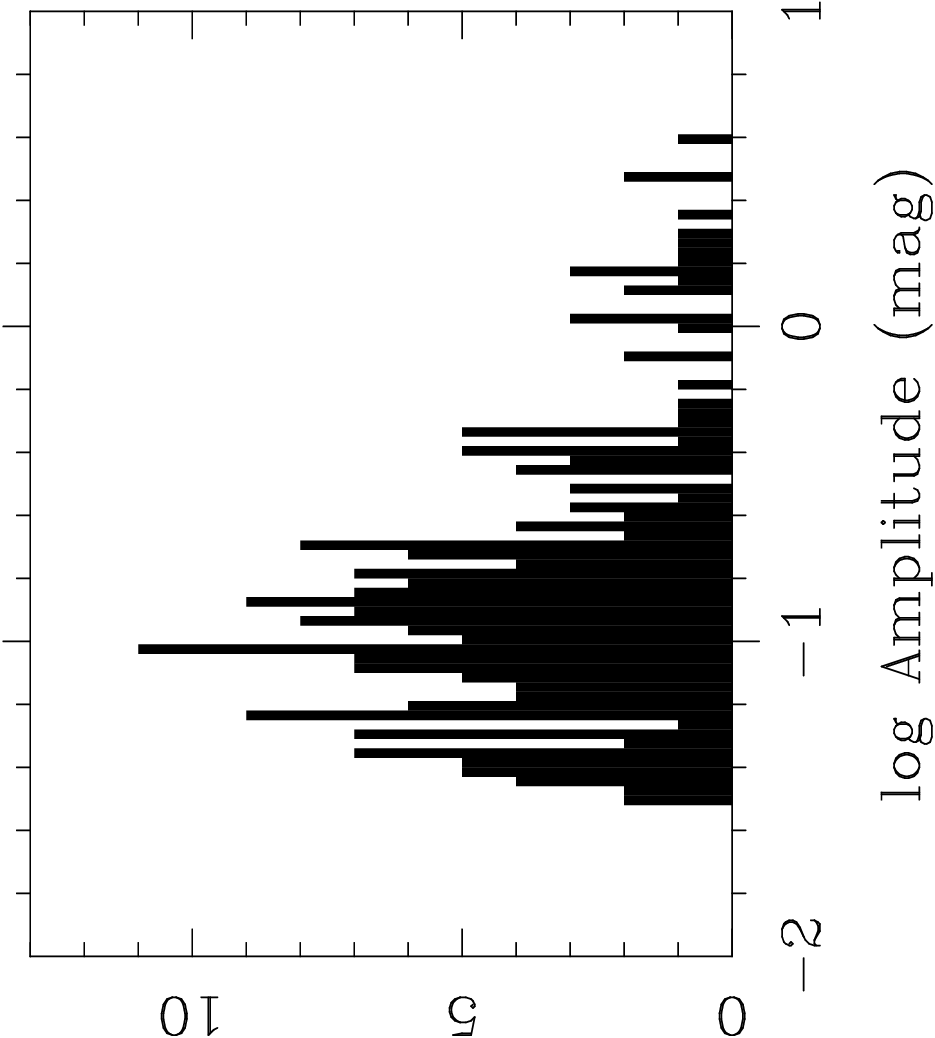}}
\noindent
\end{picture}
\vspace{98mm}
\caption{Histograms showing the cumulative period distribution (left
  panels), the period distribution (middle panels) and the amplitude
  distribution (right panels) for the short term variable point
  sources in the FSVS. Top: the 689 out of 744 sources with accuracies
  in the periods and amplitudes of the order of 30 per cent or
  less. Bottom panels: sources where the error in their periods and
  amplitudes is of the order of 10 per cent or less.}
\label{res:timeamp:fig2}
\end{figure*}

The number of variable point sources for which we can determine their
variability timescales and amplitudes by using the floating mean
periodogram test is a very good fraction of the total number of
variables detected using the $\chi^2$ test (i.e. $\sim$700 out of the
976 if we use error cutoffs of the order of 30 per cent).

Fig.~\ref{res:timeamp:fig1} shows the ratio of variable sources for
which we have determined the timescale and the amplitude of their
variability with an error less than 20 percent with the floating mean
periodogram with respect to the number of sources flagged as variable
with the $\chi^2$ test as a function of the number of observations. To
account for the fact that there are large differences between the
number of variable sources for each field, we have only plotted those
fields for which the number of variable sources is 10 or more. There
is a clear correlation between the number of observations and the
fraction of variables we can solve with the floating mean
periodogram. Including the fields with less than 10 variables
increases the scatter in the plot, whereas including only those fields
with 20 or more variables decreases the scatter. The slope of the
correlation remains the same and indicates that with a number of
observations of the order of 30, taken within a 2 week time span, we
can reconstruct most variables present in the data (with those periods
and amplitudes in the range determined in Section~\ref{an:fmp}). Of
course, this argument assumes that the lightcurves of the variables
are close to sinusoidal.

Fig.~\ref{res:timeamp:fig2} presents the distribution of variables we
find in the FSVS according to the period and amplitude of the
variability as well as their cumulative period distribution. The top
panels consider the total number of variables in the trusted range of
periods and amplitudes (689) where the error on the periods and
amplitudes is less than 30 per cent. The bottom panels present the
distributions when we only take the systems where the period and
amplitude determined has a maximum error of 10 per cent. In both cases
we find that most systems lie at short periods and low amplitudes,
with only a few systems showing larger amplitudes and periods. We find
that 50 per cent of the objects show periods below 6 hours with peaks
in the 30 per cent error distribution at $\sim$24\,min,
$\sim$0.03\,days ($\sim$43\,min), $\sim$0.12\,days ($\sim$2.9\,hours),
$\sim$0.79\,days ($\sim$19\,hours), $\sim$1.3\,days and $\sim$4\,days,
and in the 10 per cent distribution at $\sim$0.12\,days
($\sim$2.9\,hours). In the 30 per cent period distribution, the clump
of sources between $\sim$24 and 36\,min ($-$1.778$<log P<$$-$1.6)
contains 67 sources. To confirm that these are short period variables,
and not just a systematic problem caused by the sampling (after all
the minimum period we are sensitive to is 24\,min) they were inspected
by eye resulting in 80 per cent being bona-fide short period variables
with the remaining 20 per cent showing only one point off the average
brightness of the target and thus giving the short period result based
only on one point variability. These one point off sources are not
present in the 10 per cent sample.

Regarding the amplitude distribution, 50 per cent of the objects show
amplitudes lower than $\sim$0.07\,mag in the 30 per cent error sample
and lower than $\sim$0.12\,mag in the 10 per cent sample.

\begin{figure}
\begin{picture}(100,0)(-270,250)
\put(-310,240){\includegraphics{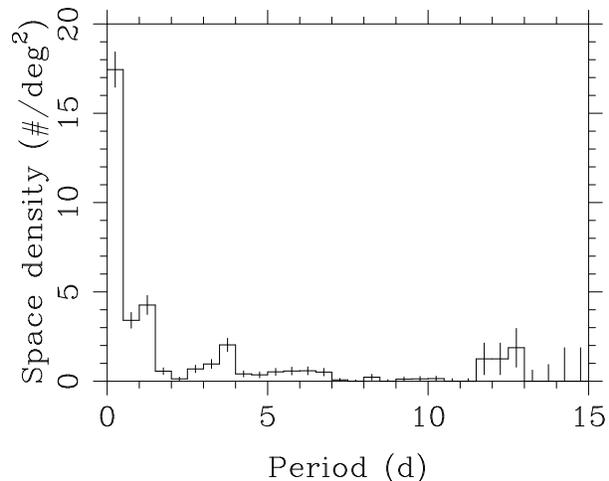}}
\noindent
\end{picture}
\vspace{60mm}
\caption{Space density of variables obtained from the FSVS.}
\label{res:timeamp:fig5}
\end{figure}

When we combine the number of sources we find per period and amplitude
bin with the sensitivity of the floating mean periodogram search,
plotted in Fig.~\ref{an:fmp:fig3}, we obtain lower limits for the
space density of variables, i.e. number of variables per square
degree, versus period. These are presented in the form of a histogram
in Fig.~\ref{res:timeamp:fig5}. We see four distinct peaks in the
distribution centred at 6 hours, 1 day, 3.75 days and 12.75 days with
a somewhat less significant peak at 6 days. The highest density of
variables show periods below 12 hours. These periods include CVs,
RR~Lyr stars, and other short period pulsators such as $\delta$\,Scuti
stars. The period range centred at 1 day includes also possible CVs,
RR~Lyr and other pulsators like $\gamma$\,Doradus stars and Pop II
Cepheids. At 3.75 days we would still find some longer period CVs,
$\gamma$\,Doradus stars, Pop II Cepheids and longer period pulsators
such as subdwarf B stars. At periods around 12.75 days, we expect to
find, apart from binaries with those orbital periods, Pop II Cepheids
contributing to the space density of variables.

\subsection{Variability colour-colour diagrams}
\label{res:ccdiag}

\begin{figure*}
\begin{picture}(100,0)(-270,250)
\put(-310,240){\includegraphics{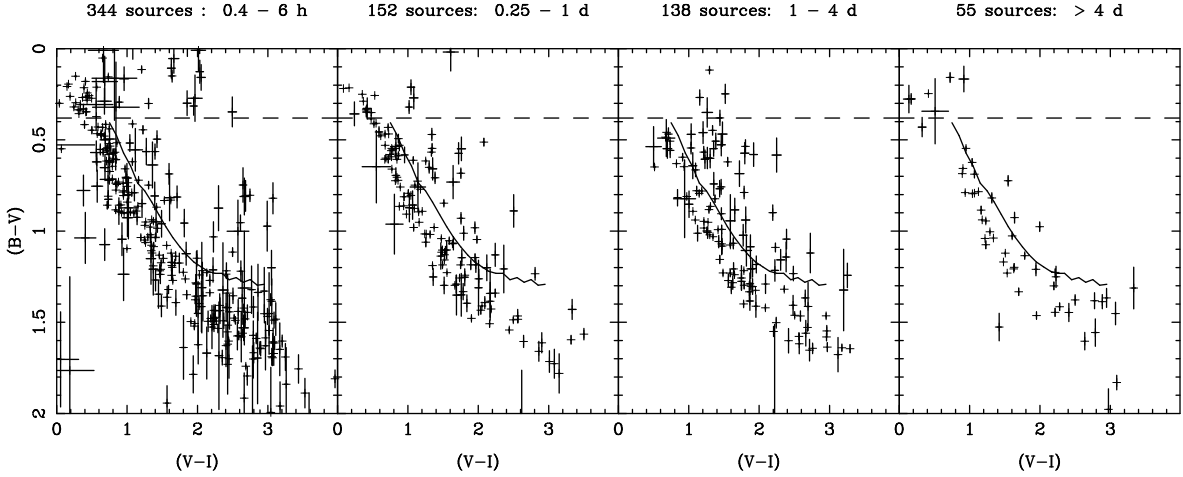}}
\noindent
\end{picture}
\vspace{65mm}
\caption{Colour-colour diagrams for several variability timescales
  obtained from the FSVS after applying a 30 per cent error cutoff in
  both the timescale and amplitude of the variability. The solid curve
  indicates the 3-$\sigma$ upper limit of the main sequence. The
  dashed line indicates the blue cut-off at (B-V)$<$0.38.}
\label{res:ccdiag:fig1}
\end{figure*}

\begin{figure*}
\begin{picture}(100,0)(-270,250)
\put(-310,240){\includegraphics{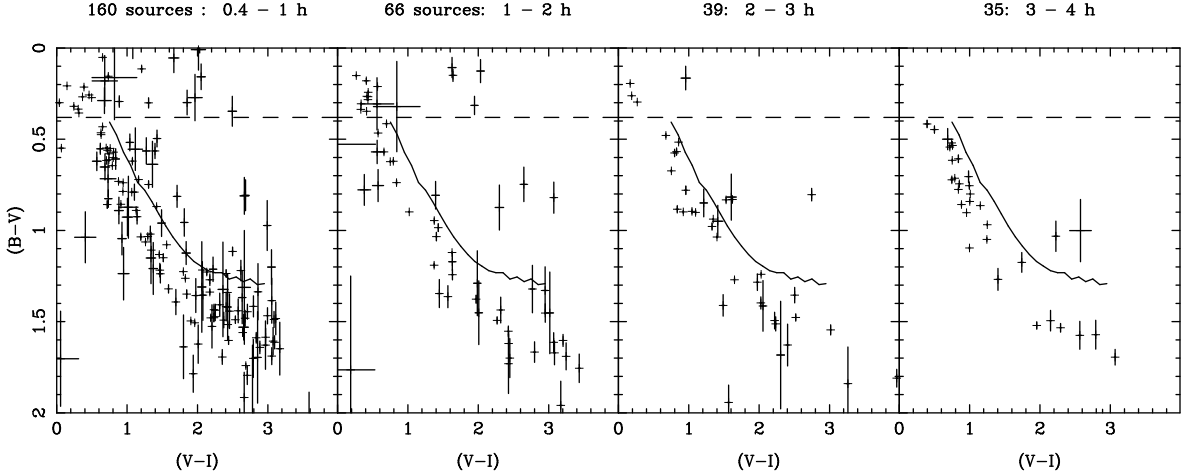}}
\noindent
\end{picture}
\vspace{65mm}
\caption{Same as in Fig.\ref{res:ccdiag:fig1} but only for short
  variability timescales. Most short period variables show variability
  timescales lower than 1\,h.}
\label{res:ccdiag:fig1b}
\end{figure*}

Keeping in mind the uncertainty of the variability timescales
determined, when we combine the variability information with the
colour information available for the FSVS we obtain the colour-colour
diagrams shown in Fig.~\ref{res:ccdiag:fig1}. We find that, if we take
only the sources with less than 30 per cent errors in their timescales
and the amplitudes, 344 point sources show variabilities shorter than
6\,h (0.25\,days). These short timescale variables are found along the
main sequence in the colour-colour diagram (see first panel of
Fig.~\ref{res:ccdiag:fig1}), where we expect to find for example
$\delta$\,Scuti stars, as well as above the main sequence and in more
extreme colour areas usually filled by binary systems where one of the
components is blue and the other red, e.g. detached red dwarf-white
dwarf binaries. At these short timescale variabilities we also find a
clump of objects above the, so called, blue cut-off at
(B$-$V)$<$0.38. The blue cut-off of the main sequence results from the
combination of the number density of different spectral types and the
scale height of the Galaxy. The colours and the short variability
timescales, of the order of characteristic close binary orbital
periods, suggest that these sources above the blue cut-off are
possibly interacting binary systems of the CV type or detached binary
systems such as subdwarf B binaries. Longer coverage, better sampled
lightcurves combined with spectroscopy are necessary in order to
identify the sources.

The variable sources with timescales shorter than 6\,h represent 50
per cent of the total number of short timescale variables in the
survey. Fig.~\ref{res:ccdiag:fig1b} shows how most of those short
period sources subdivide in smaller variability timescale ranges. We
find that about half of the short period sources have variability
timescales shorter than 1\,h. These are again distributed along the
main sequence with a few objects placed above it.

There are 152 objects showing variability in the 0.25 - 1\,d
range. These objects are also mostly distributed along the main
sequence, including $\gamma$\,Doradus pulsators amongst others, with
some cases found in the extreme colour region. The RR~Lyr variables
present in the survey should be found in this variability range. If we
combine this with the colours expected for RR~Lyr systems,
i.e. 0.1$<$(B$-$V)$<$0.45, 0.1$<$(V$-$I)$<$0.65 \cite{glw05} we find
12 RR~Lyr candidates. One of these will be discussed, as an example,
in Section~\ref{res:example}.

We find 138 sources that show variabilities between 1 and 4\,d again
distributed mostly along the main sequence. This would include
$\gamma$\, Doradus pulsators as well as Pop II Cepheids. There are 55
point sources that show variabilities on timescales longer than
4\,d. Binary systems with these periods as well as Pop II Cepheids are
included in this period range. The blue sources found in these two
period ranges above the blue cutoff could be subdwarf B slow pulsators
or binaries.

\begin{figure*}
\begin{picture}(100,0)(-270,250)
\put(-310,240){\includegraphics{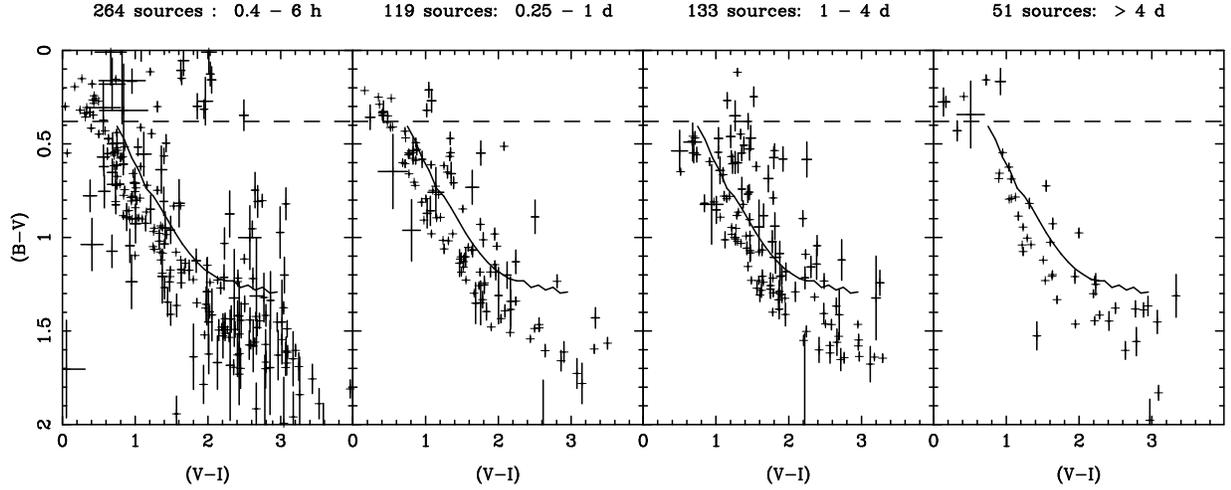}}
\noindent
\end{picture}
\vspace{65mm}
\caption{Same as in Fig.\ref{res:ccdiag:fig1} but with error limits
  set to 20 for both timescales and amplitudes.}
\label{res:ccdiag:fig2}
\end{figure*}

Fig.~\ref{res:ccdiag:fig2} presents similar diagrams to those in
Fig.~\ref{res:ccdiag:fig1} but for the sources where the error limits
were set to 20 percent. The distribution of variables in the diagrams
is very similar to the initial one. Again about half of the sources
show variability timescales shorter than 6\,h. Most of the objects
that have disappeared from the diagram come from the shorter period
ranges (23 per cent out of the 0.4 - 6\,h bin and 22 per cent out of
the 0.25 - 1\,d bin).  This could be due to the fact that if the
signal is not sinusoidal or regular, the floating mean periodogram
tends to calculate periods shorter than the input ones with large
errors. It is worth noticing that a few of the interesting objects
above the blue cut-off have disappeared. 

\subsection{Fraction of variable sources as a function of spectral type}
\label{res:varnovar:mainseq}

The fraction of variable point sources along the main sequence, as a
function of spectral type (as defined in Johnson 1966), is presented
in Table~\ref{res:varnovar:mainseq:tab1}. The number of variables used
here is that obtained from the $\chi^2$ test. The fraction of
variables is constant with spectral type and has a value around 1 per
cent.

We also present the distribution of variability periods and amplitudes
measured for the number of variables that were analysed with the
floating mean periodogram algorithm. Four period bins and four
amplitude bins are presented in
Table~\ref{res:varnovar:mainseq:tab2}. Some spectral type ranges
contain very few objects. For those with a larger number of objects
(K0 to M5) the variable sources spread themselves in similar fractions
in the spectra ranges K0 to M0 with more short period systems in the
spectral range M0 -- M5. The fraction of variables with timescales
longer than 4 days is significantly smaller than in the other
timescale bins. In the case of the amplitudes of the variability, most
sources show variabilities with amplitudes lower than 0.1 mags.

\begin{table*}
\caption{Fraction of variable sources per spectral type bin found in
  the FSVS. Notice that although the first main sequence bin (spectral
  types F0--F5) contains more variables than the rest, this is likely
  to be the result of including non main sequence sources in that
  bin. Non MS stands for non main sequence sources. Two fractions are
  given, for all sources in the FSVS and only for those with more than
  3 V measurements.}
\label{res:varnovar:mainseq:tab1}
\begin{center}
\begin{tabular}{cccccc}
\hline
Spectral type & B$-$V & V$-$I & \# Var & \# Sources & Fraction (\%) \\
\hline
non MS    & $<$ 0.38   & 0.47--0.64 & 11 & 700 & 1.6 \\
non MS    & $<$ 0.38   & 0.00--0.4 & 29 & 833 & 3.5 \\
F0--F5 & 0.31--0.43 & 0.47--0.64 & 13 & 602 & 2.2 \\
F5--G0 & 0.43--0.59 & 0.64--0.81 & 62 & 6054 & 1.0 \\
G0--G5 & 0.59--0.66 & 0.81--0.89 & 20 & 1816 & 1.1 \\
G5--K0 & 0.66--0.82 & 0.89--1.06 & 63 & 6814 & 0.9  \\
K0--K5 & 0.82--1.15 & 1.06--1.62 & 204 & 20207 & 1.0 \\
K5--M0 & 1.15--1.41 & 1.62--2.19 & 133 & 12974 & 1.0 \\
M0--M5 & 1.41--1.61 & 2.19--3.47 & 153 & 16256 & 0.95 \\
M5--M8 & 1.61--2.00 & 3.47--4.70 & 3 & 179 & 1.7 \\
\hline
\end{tabular}
\end{center}
\end{table*}

\begin{table*}
\caption{Period and amplitude distribution of the main sequence
  variables analysed with the floating mean periodogram. Four bins for
  the variability timescale and four for the variability amplitude are
  shown. The percentage of sources in each bin is given. FMP stands for
  floating mean periodogram. Amplitudes are given in magnitudes.}
\label{res:varnovar:mainseq:tab2}
\begin{center}
\begin{tabular}{cc|rccl|rccl}
\hline
Spectral type & \# VarFMP & \multicolumn{4}{c}{Period distribution(\%)} &
\multicolumn{4}{c}{Amplitude distribution(\%)}\\
 & & 0.4--0.6\,h & 0.25--1.0\,d & 1.0--4.0\,d & $>$4\,d &
0.01--0.1 & 0.1--0.25 & 0.25--1.0 & $>$1.0\\
\hline
F0--F5 & 6  & 50.0 & 33.3  & 0  & 16.6  & 16.6   & 50.0  & 16.6 & 16.6 \\ 
F5--G0 & 30 & 53.3 & 23.3  & 23.3  & 0  & 60  & 26.6  & 10.0 & 3.3 \\ 
G0--G5 & 6   & 50.0 & 33.3  & 16.6  & 0  & 83.3   & 16.6  & 0 & 0 \\ 
G5--K0 & 18  & 61.1 & 22.2  & 5.5 & 11.1  & 88.8  & 11.1  & 0 & 0 \\ 
K0--K5 & 86  & 43.0 & 23.2 & 23.2 & 10.5 & 82.6 & 11.6 & 5.8 & 0 \\ 
K5--M0 & 66  & 36.4 & 33.3 & 25.8 & 4.5  & 77.3  & 12.1 & 7.6 & 3.0 \\ 
M0--M5 & 69  & 63.8 & 10.1 & 18.8 & 7.2  & 78.3  & 17.4 & 4.3 & 0 \\ 
M5--M8 & 2   & 100.0  & 0  & 0  & 0  & 50   & 50  & 0 & 0 \\ 
\hline
\end{tabular}
\end{center}
\end{table*}

\subsection{Example lightcurves}
\label{res:example}

\begin{figure*}
\begin{picture}(100,0)(-270,250)
\put(-310,240){\includegraphics{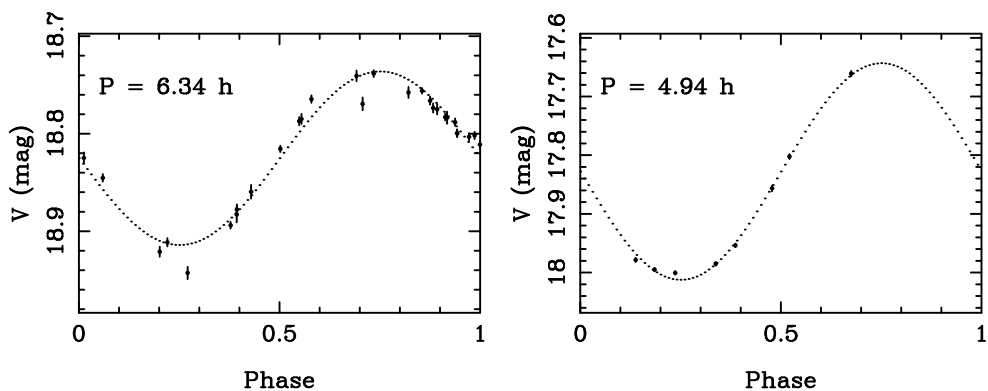}}
\noindent
\end{picture}
\vspace{55mm}
\caption{Lightcurves of two example objects folded on the period
  determined with the floating mean periodogram. The left hand side
  lightcurve corresponds to one of the RR~Lyr candidates discussed in
  Section~\ref{res:ccdiag} and the right hand side lightcurve to a
  subdwarf B star candidate. The best sinusoidal fit is also plotted.}
\label{res:example:fig1}
\end{figure*}

Fig.~\ref{res:example:fig1} presents the lightcurves for two example
sources found in our data. The examples have only been selected to
illustrate two very different samplings. The first example corresponds
to one of the candidate RR~Lyr (see Section~\ref{res:ccdiag}) for
which we determined a period of 6.34\,h from the floating mean
periodogram. The second example corresponds to a source with very blue
colours, for which we have obtained spectra that suggests it is a
subdwarf B star. The period measured for this system, which probably
corresponds to its orbital period in a binary system, is 4.95\,h. For
both sources we present the measurements as a function of phase folded
at the period determined from the floating mean periodogram. The best
sinusoidal fit to the data is also plotted.

The FSVS covers a region in the sky in which there are two known
cataclysmic variables, GO~Com and V394~Lyr. These two sources were
identified as variables in both the $\chi^2$ and the floating mean
periodogram tests. Interestingly enough, both of them went into
outburst during the observations which made it more difficult to
determine their orbital periods. GO~Com was observed in three
different epochs and it rose by a magnitude in V from the second to
the third epoch (from 19.4 to 18.5). In an attempt to subtract the
outburst contribution to the variability, and thus to be able to
measure its orbital period, we calculated its average brightness for
each epoch and subtracted the average from the measurements taken on
that epoch. When we run the floating mean periodogram on the resulting
lightcurve we obtain a periodogram with two main groups of aliases,
one centred in 30\,min and another on 90\,min. The sampling of the
lightcurve does not allow us to determine the period with more
accuracy than this. The second alias lies near the 94.8\,min period
determined by Howell et al. \shortcite{hsc95}. V394~Lyr was observed
in 7 epochs and it rose by two magnitudes in V from the first to the
sixth (from 19.1 to 17) starting to decrease on the the seventh. We
followed the same steps outlined above for GO~Com and found the lower
$\chi^2$ aliases to be 43\,min and 4.32\,h. There are no measurements
of the orbital period of V394~Lyr in the literature that we can
compare these values with. The folded lightcurves on these two periods
do not look convincing which lead us to think that none of these are
the true orbital period of the system.

\subsection{Sources not detected in all three bands}
\label{res:extremecolour}

A third of the variable point sources found in the FSVS using the
$\chi^2$ test were not detected either in B, I or in both B and I
bands. In a handful of cases this was due to the point source being
close to the CCD boundary, in 21 percent of the cases the source
appeared as blended with another one, in 28 per cent of the cases the
source was not detected because it was too faint, and in 50 per cent
of the cases the source was saturated in the I band. Most of these
sources saturated in the I band are very bright in the three bands (71
percent) indicating that about one sixth of the variable sources found
are at the bright end of the survey. This is expected as most
variables probably show low amplitude variations which are easier to
detect for bright sources. In the case of the blended sources,
blending would most probably affect the V band magnitudes perhaps
introducing spurious variability. 

On the other hand, 29 percent of the sources undetected in B and/or I,
do show larger colour differences, (B$-$I)$>$3 and up to 5 magnitudes
in some cases. Colour differences of (B$-$I)$>$3 are expected for some
main sequence stars and do not imply that we are dealing with extreme
colour objects. The lower limit colours calculated for these sources
(by assuming their I band magnitude is equivalent to the I band
magnitude of the brightest unsaturated star in the field) fill an area
of the colour space, which is also filled by the point sources that
were detected (unsaturated) in the three bands. This indicates that,
although their colours are more extreme, they are not unusual compared
with the sources detected in all three bands, they are just brighter
and therefore saturated in the I band.

The same analysis carried out in Section~\ref{res:timeamp} for point
sources with B, I and V detections, and more than 4 V measurements can
be carried out for 154 sources with no B and/or I detections. In these
cases the information we have of the variable sources is only their
timescales and amplitudes and not their colours. Once we discard
sources for which their calculated periods and amplitudes lie outside
the trustable ranges determined in Section~\ref{an:fmp} for each
field, we are left with 71 short period variable sources.  We find
that 29.5 per cent of the variables with 30 per cent accuracy in their
periods and amplitudes show periods between 0.4 and 6 hours, 39.3 per
cent between 6 hours and 1 day, 21.3 per cent between 1 and 4 days,
and 9.8 per cent of more than 4 days. These values are slightly
different (19, 23.8, 33.3 and 23.8 respectively) if we consider the 21
sources where the periods and amplitudes determined have errors of
less than 10 per cent. When we compare this distribution of variables
in period bins with the one we found for systems with B and I
measurements (50, 22 20, 8 per cent respectively for each period bin),
we find that the number of shorter period variables (first period
range) is smaller for the sources with no B and/or I detections and
the number of variables in the last two period bins is larger.

\section{Conclusions}

We have analysed the short timescale variability information contained
in the FSVS and find that about 1 per cent of all point sources are
variable. Of those variables, about 50 per cent show variability
timescales shorter than 6 hours, 22 per cent show variabilities
between 6 hours and 1 day, 20 per cent between 1 and 4 days and 8 per
cent show periods longer than 4 days. The distribution of variables
with spectral type is fairly constant along the main sequence, with 1
per cent of all the sources being variable, except at the blue end of
the main sequence where the fraction of variable sources increases
possibly due to contamination by non main sequence sources. Above the
main sequence, beyond the blue cut-off at (B$-$V)$<$0.38, we find that
the fraction of variables increases to 3.5 percent.

The highest space density of variables found in the FSVS (i.e. 17 per
deg$^2$) show periods below 12 hours. These include CVs, RR~Lyr stars,
and other short period pulsators such as $\delta$\,Scuti stars. We
find a density of 4 variables per deg$^2$ centred at a 1 day period
which includes longer period CVs, RR~Lyr and other pulsators like
$\gamma$\,Doradus stars and Pop II Cepheids. A space density of 2
variables per deg$^2$ at 3.75 days includes, some longer period CVs,
$\gamma$\,Doradus stars, Pop II Cepheids and longer period pulsators
such as subdwarf B stars. At 12.75 days we also find 2 variables per
deg$^2$. These would be mainly binaries with those orbital periods and
Pop II Cepheids.

It is easier to compare these space densities with those expected for
the mentioned populations when we combine the period information with
the colours of the populations under study. The case of CVs and many
pulsators is complicated as they appear mixed through several period
and colour ranges and in many cases it is necessary to obtain spectra
to confirm the nature of the variable source. The space densities of
CVs and subdwarf B stars will be studied in detail in a future
paper. In the case of RR Lyr stars, we find 3 certain members and 9
other candidates down to V = 21.6. Assuming we have detected all RR
Lyr between V = 16--22, we determine a space density of
$\sim$10$^{-3}$kpc$^{-3}$ in agreement with the space density
determined by Preston, Shectman \&\ Beers \shortcite{psb91} at a
distance of 100--150kpc from the Galactic Centre.

By using the floating mean periodogram, we have determined the most
likely periods and amplitudes of a fraction of the variables found in
the FSVS. We find that we are complete down to V = 22 for CVs in the
minimum period (80 min) as long as they show variability amplitudes of
the order of 0.4 mag. We are complete down to V = 22 for periods
between 80 min and 1 day in a 17.82\,deg$^2$ area of the survey as
long as the amplitude of the variability is at least 0.7 mag. This
includes most RR Lyr stars. We will be able to detect RR Lyr also down
to V = 23 when their variability amplitudes are at least 1.5 mag.

\section*{Acknowledgements}

We thank T. R. Marsh for making his analysis software available. The
FSVS was supported by NWO Spinoza grant 08-0 to E. van den Heuvel. The
FSVS is part of the INT Wide Field Survey. LM-R, PJG and EvdB are
supported by NWO-VIDI grant 39.042.201 to PJG. GN is supported by
NWO-VENI grant 639.041.405. The Isaac Newton telescope is operated on
the island of La Palma by the Isaac Newton Group in the Spanish
Observatorio del Roque de los Muchachos of the Instituto de
Astrof\'{\i}sica de Canarias.

\end{document}